\documentclass[superscriptaddress, twocolumn]{revtex4-1}
\usepackage{amsmath,latexsym,amsfonts,amssymb}
\usepackage{graphicx}
\usepackage{inputenc}
\usepackage{color}
\usepackage{setspace}
\usepackage{array}
\usepackage{tabularx}
\usepackage{natbib}
\usepackage{engrec}

\newcommand{\ndzr}{Nd$_2$Zr$_2$O$_7$}
\newcommand{\hoti}{Ho$_2$Ti$_2$O$_7$}
\newcommand{\tbti}{Tb$_2$Ti$_2$O$_7$}
\newcommand{\erti}{Er$_2$Ti$_2$O$_7$}
\newcommand{\ybti}{Yb$_2$Ti$_2$O$_7$}

\newcommand{\nd}{Nd$^{3+}$}

\begin{document}

\author{E. Lhotel}
\email[]{elsa.lhotel@neel.cnrs.fr}
\affiliation{Institut N\'eel, CNRS and Universit\'e Grenoble Alpes, 38042 Grenoble, France}
\author{S. Petit}
\email[]{sylvain.petit@cea.fr}
\affiliation{Laboratoire L\'eon Brillouin, CEA CNRS  UniversitŽ Paris Saclay, CE-Saclay, F-91191 Gif-sur-Yvette, France}
\author{M. Ciomaga Hatnean}
\affiliation{Department of Physics, University of Warwick, Coventry, CV4 7AL, United Kingdom}
\author{J. Ollivier}
\affiliation{Institut Laue Langevin, F-38042 Grenoble, France}
\author{H. Mutka}
\affiliation{Institut Laue Langevin, F-38042 Grenoble, France}
\author{E. Ressouche}
\affiliation{INAC, CEA and Universit\'e Grenoble Alpes, CEA Grenoble, F-38054 Grenoble, France}
\author{M. R. Lees}
\affiliation{Department of Physics, University of Warwick, Coventry, CV4 7AL, United Kingdom}
\author{G. Balakrishnan}
\affiliation{Department of Physics, University of Warwick, Coventry, CV4 7AL, United Kingdom}

\title{Evidence for dynamic kagome ice}

\begin{abstract} 
The search for two-dimensional quantum spin liquids, exotic magnetic states remaining disordered down to zero temperature, has been a great challenge in frustrated magnetism over the last few decades. Recently, evidence for fractionalized excitations, called spinons, emerging from these states has been observed in kagome and triangular antiferromagnets. In contrast, quantum ferromagnetic spin liquids in two dimensions, namely quantum kagome ices, have been less investigated, yet their classical counterparts exhibit amazing properties, magnetic monopole crystals as well as magnetic fragmentation. Here we show that applying a magnetic field to the pyrochlore oxide \ndzr, which has been shown to develop three-dimensional quantum magnetic fragmentation in zero field, results in a dimensional reduction, creating a dynamic kagome ice state: the spin excitation spectrum determined by neutron scattering encompasses a flat mode with a six arm shape akin to the kagome ice structure factor, from which dispersive branches emerge.

\end{abstract}

\maketitle

\section{Introduction}

\begin{figure}[t]
\includegraphics[width=9cm]{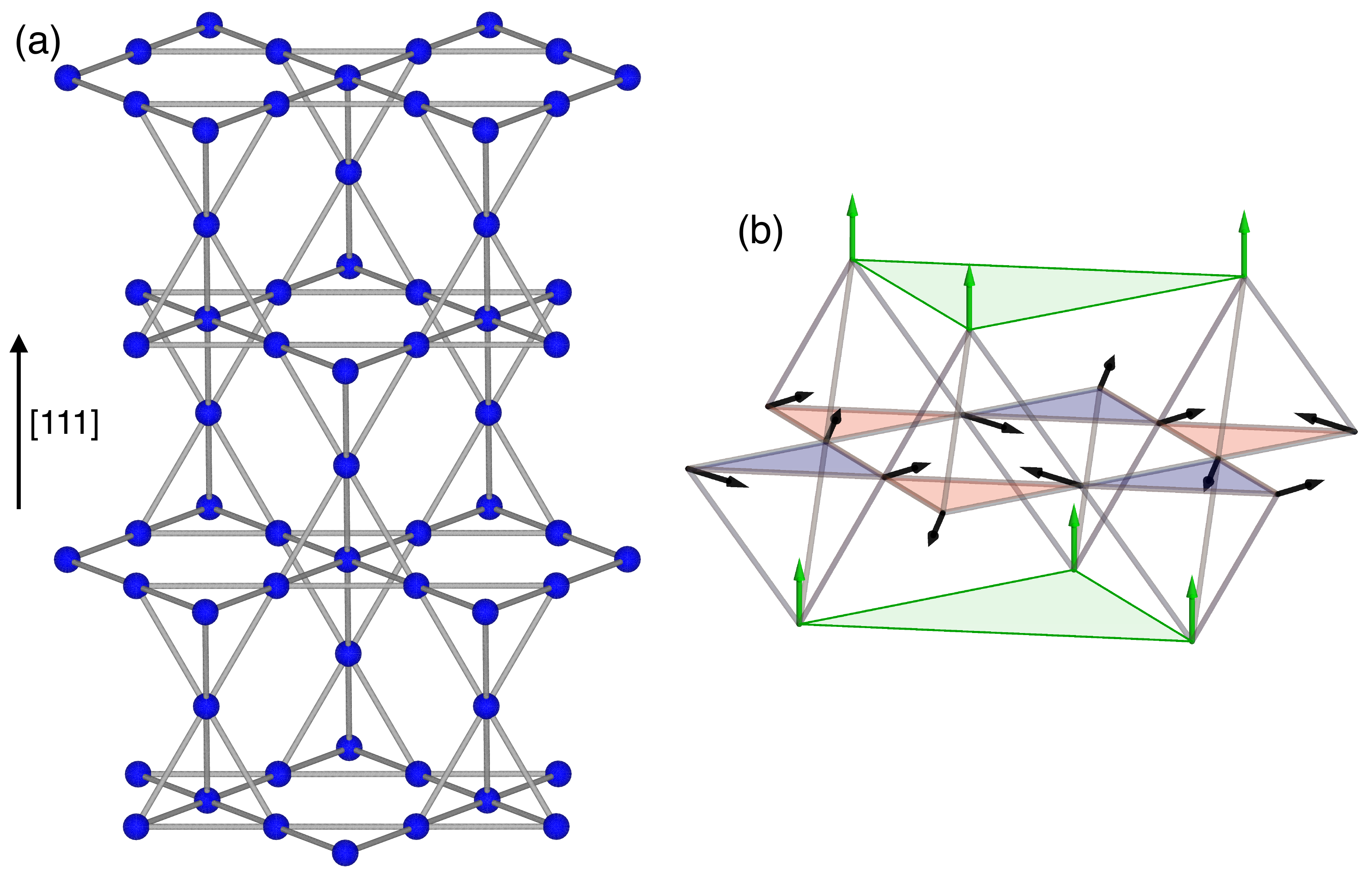}
\caption{\label{fig_kagome} {\bf Kagome ice state in the pyrochlore lattice.} (a) View of the pyrochlore structure along the $[111]$ direction showing the stacking of triangular and kagome planes (See also Supplementary Table 1 and Supplementary Figure 1). (b) Classical kagome ice state when a magnetic field is applied along [111]. The apical spins in the triangular planes (green triangles) are parallel to the field, while the kagome spins obey the ice-rule 2 in -- 1 out (red triangles) / 1 in -- 2 out (blue triangles). }
\end{figure}

The two-dimensional kagome and three-dimensional pyrochlore structures are low connectivity lattices based on corner sharing triangles or tetrahedra respectively. They form a rich playground to study unconventional magnetic states, such as spin liquids, induced by geometrical frustration. At first glance, they bear no relation to each other, especially when considering their dimensionnality. Nevertheless, along $[111]$ (and symmetry related directions), the pyrochlore lattice can be viewed as a stacking of triangular and kagome layers, as illustrated in Figure \ref{fig_kagome}(a). As a result, if one is able to decouple these layers, two-dimensional physics characteristic of the kagome lattice can develop on the pyrochlore lattice.

This is of specific interest in the context of spin-ice, an original state of matter made of an assembly of Ising spins aligned along the local $\langle 111 \rangle$ directions (which join at the centre of the tetrahedra) and coupled by a ferromagnetic interaction \cite{Harris97}. Spin-ice is a degenerate state where the spins locally obey, on each tetrahedron, the 2 in -- 2 out ice-rule, i.e. two spins point in and two spins point out of each tetrahedron. Two-dimensional kagome physics is observed in spin-ice when a magnetic field is applied along the $[111]$ direction: the spins in the triangular planes having their easy axis parallel to the field, they are easily polarized, and thus decouple from the kagome layers. Provided the field is not too strong, a degeneracy persists within the kagome planes, characterized by the kagome ice-rule \cite{Moessner03}, i.e. 2 spins point into each triangle, and 1 out, or vice versa, as shown in Figure \ref{fig_kagome}(b). 
This corresponds to the two-dimensional kagome ice state, extensively studied in artificial lattices \cite{Nisoli13, Canals16} and recently realized in a bulk material \cite{Paddison16}. 
In this state, the algebraic correlations within the tetrahedra characteristic of the spin ice state \cite{Isakov04, Henley05} become two-dimensional within the kagome planes \cite{Moller09, Chern11}. They give rise in both cases to a diffuse neutron scattering signal exhibiting pinch points, but with different patterns \cite{Fennell09, Tabata06}. This field induced 3D-2D reduction also manifests itself as a magnetization plateau at $2/3$ of the saturation magnetization \cite{Sakakibara03}. 

\begin{figure}[h!]
\includegraphics[width=7.5cm]{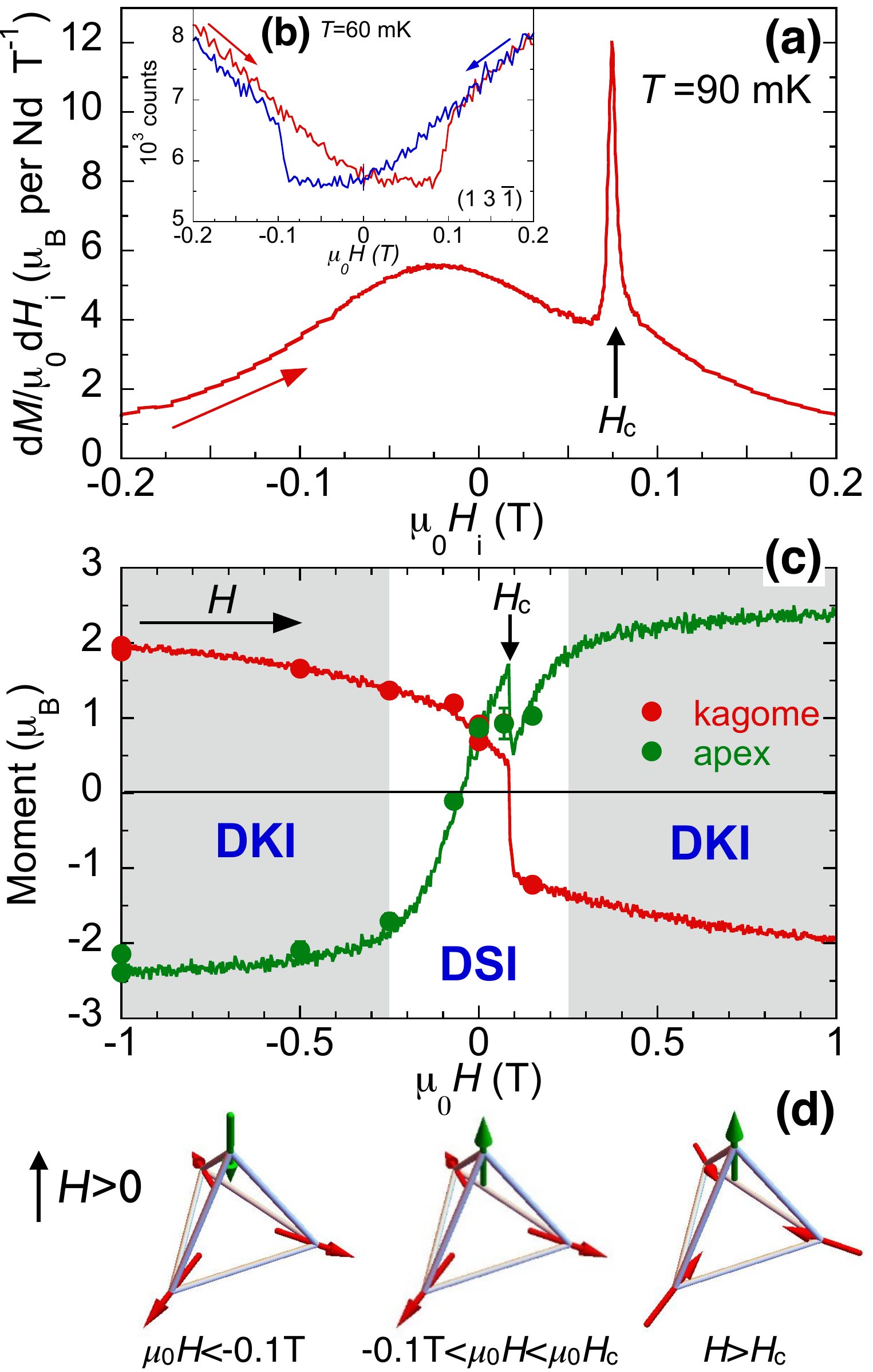}
\caption{\label{fig_struct} {\bf Field evolution of the magnetic structure for ${\bf H \parallel [111]}$.} (a) Derivative of the magnetization $dM/dH_i$ vs the internal magnetic field $H_i$ when $H$ is swept from negative to positive. (b) Field dependence of the $(1 3 \bar{1})$ Bragg peak intensity when sweeping the field from negative to positive (red) and back (blue). (c) Refined magnetic moment for the apex (green) and kagome (red) spins of a tetrahedron as a function of field. Points correspond to data collections. The error bars are provided by the Rietveld refinement made using Fullprof.  Lines are obtained from the analysis of the field sweeping measurements where the moment values have been rescaled to the refined values obtained from the data (see Supplementary Figures 2-4 and Supplementary Note 1). The obtained moments are aligned along the $\langle 111 \rangle$ directions. By convention, for an up tetrahedron, a positive (negative) moment value corresponds to an out (in) moment. The regions where a dynamic kagome ice state (DKI) is stabilized are coloured in grey. At low field, the dynamic spin ice state (DSI) is observed. (d) Schematic of the magnetic structure of the three observed configurations for an up tetrahedron: (i) 1 in -- 3 out for $\mu_0 H<-0.1$ T, (ii) all out around $H=0$, (iii) 3 in -- 1 out for $H>H_c$. Note that only the spin directions are meaningful and not their size: apart from at saturation and at $H=0$, apex and kagome spins do not carry the same ordered moment. 
}
\end{figure}

The way this classical picture is affected by quantum effects, and especially the conditions under which a quantum kagome ice state could be stabilized from a quantum spin ice state has aroused great interest \cite{Carrasquilla15, Owerre16, Onoda17}. For instance, this issue has been tackled in quantum spin-ice candidates like \tbti \cite{Molavian09, Yin13, Takatsu16} and the Pr pyrochlores \cite{Machida10, Sibille16}, through the search for magnetization plateau. In this work, we address the kagome ice physics in a different way by focusing on the effect of a $[111]$ field on a dynamic spin ice state. Our starting point is the peculiar behaviour of the \ndzr\, quantum pyrochlore magnet, where classical spin ice physics is considerably modified by the existence of transverse terms in the Hamiltonian: in zero field, the ground state exhibits an all in -- all out magnetic structure, while spin ice signatures are transferred in the excitation spectrum, taking the form of a flat spin ice mode \cite{Petit16}. Applying a $[111]$ magnetic field, we show that the flat mode persists and that a dimensional reduction occurs in the excitation spectrum: above about 0.25 T, a flat kagome ice like mode forms in the excitation spectrum, featuring a dynamic kagome ice state. Mean-field calculations using the XYZ Hamiltonian \cite{Huang14} adapted for the \nd\ ion allow us to refine a set of exchange parameters. Some discrepancies between our observations and the calculations, however, point to the existence of more complex processes.

\section{Results}

{\bf Field induced magnetic structures}\\

The $[111]$ field induced phase diagram in \ndzr\ has been probed by magnetization measurements \cite{Lhotel15, Opherden17}. A small anomaly attributed to a magnetic transition is observed at $\mu_0H_c \approx 0.08$ T, with a hysteretic behaviour, on top of a smooth evolution which is not expected for conventional Ising spins. Bragg peak intensities measured by neutron diffraction also show a hysteretic behaviour and a discontinuity at $H_c$, confirming that the cusp in the derivative of the magnetization $dM/dH$ corresponds to a change in the magnetic structure (see Figures \ref{fig_struct}(a) and (b)). The value and orientation of the magnetic moments obtained from the magnetic structure refinements (see Supplementary Note 1) are shown in Figure \ref{fig_struct}(c) and (d). Over the whole field range, the spins lying in the kagome planes adopt a 3 in -- 3 out configuration. In a large enough magnetic field, typically 1 T, both types of spins, i.e. the apex and the kagome spins, are saturated, forming the expected ordered classical structure 3 in -- 1 out / 1 in -- 3 out. Starting from $-1$ T and sweeping the field up, the ordered components progressively decrease. The apical spin totally loses its ordered moment at about $- 0.1$~T, before flipping to the zero field configuration, an all in -- all out structure with a partially ordered moment of $0.8~\mu_{\rm B}$ (to be compared to $2.3~\mu_{\rm B}$, the magnetic moment of the ground doublet) \cite{Lhotel15}. When further increasing the field, the kagome spins flip at $H_c$ to accommodate the field, and the system returns to a 3 in -- 1 out / 1 in -- 3 out structure. Finally, at larger fields, the ordered magnetic moments continue increasing towards the saturated value.

\begin{figure*}[!t]
\includegraphics[width=\textwidth]{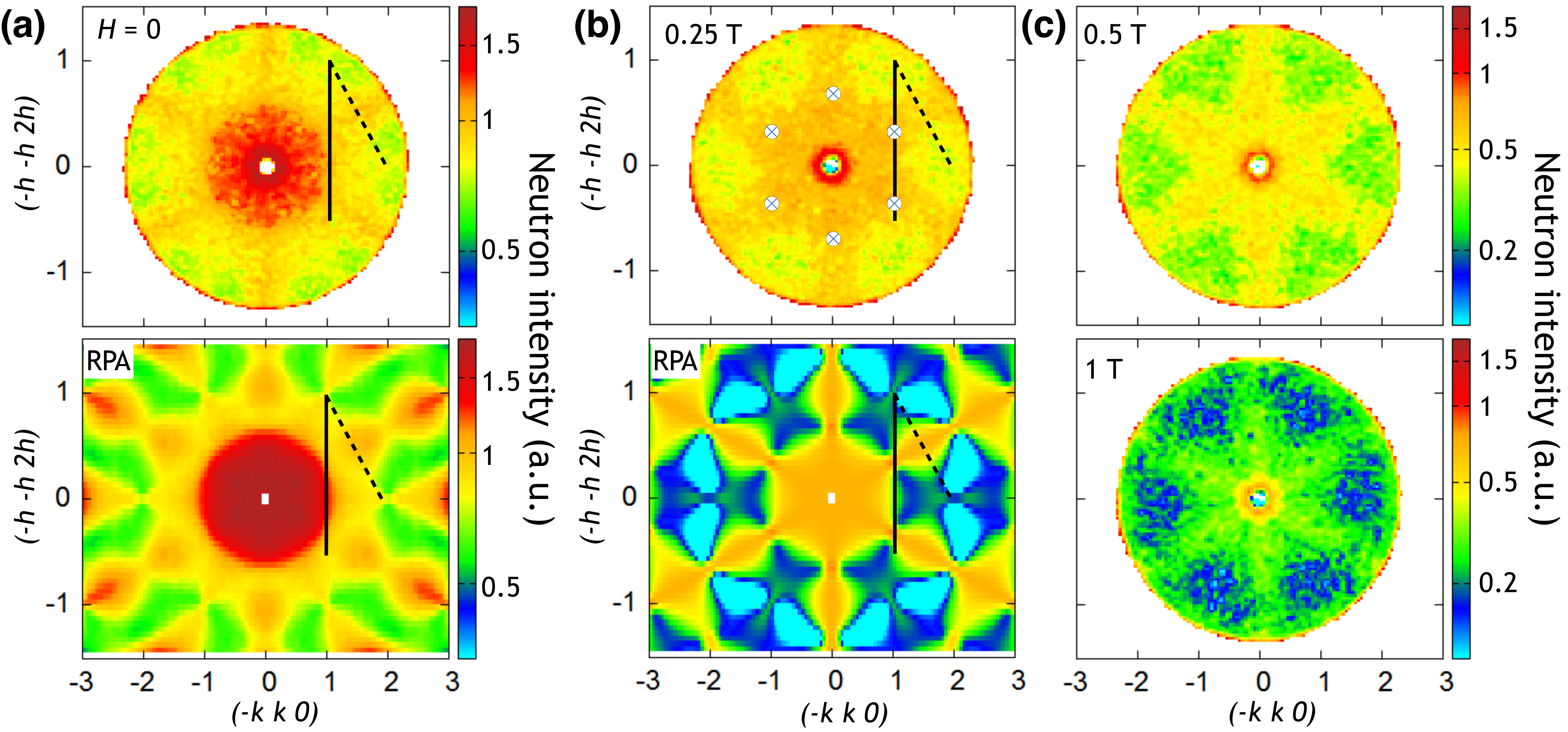}
\caption{\label{fig_INS} {\bf Dynamic kagome ice state seen in inelastic neutron scattering for ${\bf H \parallel [111]}$.} Inelastic intensity averaged around $E=(50 \pm 5)~\mu$eV (corresponding to the energy shown by the red arrow in Figure \ref{fig_INS_calc}). (a) Zero-field measurements at 60~mK and Random Field Approximation (RPA) calculations for the pseudo-spin 1/2 model described in the text with ${\sf J}_x= -0.36$~K, ${\sf J}_y=0.066$~K, ${\sf J}_z=0.86$~K and ${\sf J}_{xz}= 0.44$~K at $T=0$. (b) Measurements in 0.25 T at 60 mK and calculations with the same parameters. The crossed circles mark the position of the expected pinch points, which appear blurred in the measurements. (c) Measurements at 0.5 T and 1 T. The black full and dashed lines indicate the directions of the slices along $(-1-h, 1-h, 2h)$ and $(-2, h, 2-h)$ shown in Figure \ref{fig_INS_calc}.
}
\end{figure*}

These field induced magnetic structures qualitatively agree with the conventional behaviour of an all in -- all out system in a $[111]$ magnetic field. Nevertheless, only a fraction of the expected Nd moment is involved in the ordered magnetic moment in the low-field region, and the magnetization increases smoothly. This is due to the peculiar dipolar octupolar nature of the ground-state Nd doublet \cite{Huang14}, which makes the magnetic moment different from a classical Ising spin and allows for non-magnetic transverse components. This results in exotic dynamics, that we have probed by inelastic neutron scattering measurements.\\

{\bf Evidence for a dynamic kagome ice mode}\\

In zero field, as previously mentioned, a dynamic spin ice mode is observed \cite{Petit16} at an energy of about 70 $\mu$eV. In the scattering plane perpendicular to $[111]$, this mode is characterized by the star-like pattern shown in Figure \ref{fig_INS}(a), with a strong intensity around ${\bf q}={\bf 0}$. 
On increasing the field, one could expect this feature to disappear at $H_c$. However, the star-like pattern persists up to 0.25 T, where it changes into a pattern with a new structure, at about the same energy, as shown in Figure \ref{fig_INS}(b): six arms appear, while the scattering intensity decreases around ${\bf q}={\bf 0}$ and the $(2,-2,0)$ ${\bf q}$-vectors. Upon increasing the field, the intensity of the arms decreases, but clearly persists up to at least 1~T (see Figure \ref{fig_INS}(c)). 

\begin{figure*}[!t]
\includegraphics[width=\textwidth]{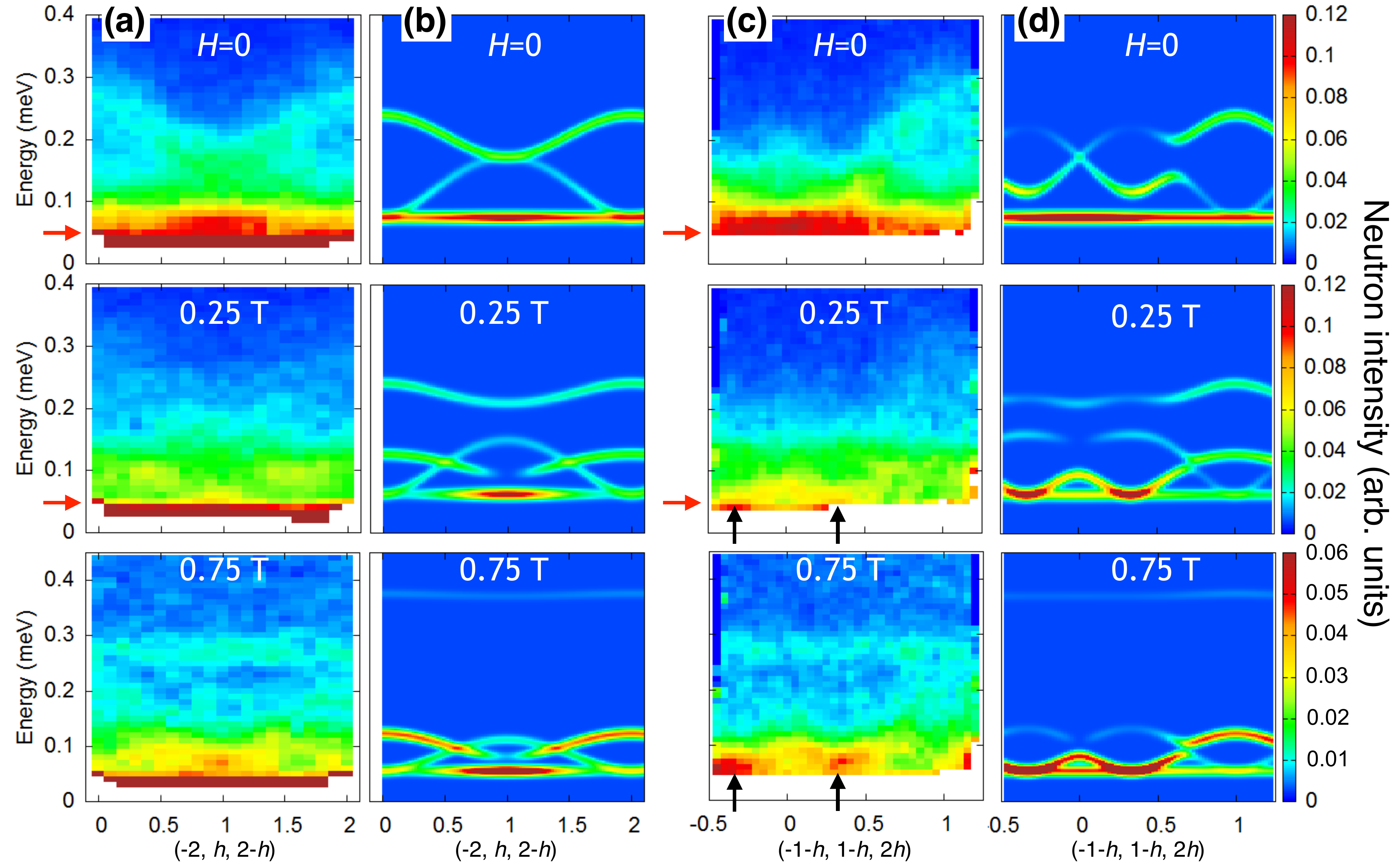}
\caption{\label{fig_INS_calc} {\bf Magnetic excitation spectra. } Slices along $(-2,h, 2-h)$ (dashed line in Figure \ref{fig_INS}) in zero field, 0.25 T and 0.75 T: (a) Measurements at 60 mK. (b) RPA calculations with ${\sf J}_x= -0.36$~K, ${\sf J}_y=0.066$~K, ${\sf J}_z=0.86 $~K and ${\sf J}_{xz}= 0.44$~K at $T=0$. Slices along $(-1-h, 1-h, 2h)$ (black line in Figure \ref{fig_INS}) in zero field, 0.25 T and 0.75 T: (c) Measurements at 60 mK. (d) RPA calculations with the same parameters. The black arrows mark the positions of the kagome ice pinch points, from which the dispersive branches emerge. The red arrows mark the position of the energy cuts of Figure \ref{fig_INS}, slightly below the actual position of the mode to minimize the integration of the dispersive branch. The colour scale at 0.75 T has been changed to emphasize the flat band at 0.26 meV in the measurements and 0.36 meV in the calculations.
}
\end{figure*}

The obtained pattern actually resembles the kagome ice neutron scattering function. Nevertheless, the pinch points expected at ${\bf q}=(2/3,2/3,-4/3)$ (and related symmetry positions), and characteristic of the existence of algebraic correlations, are not clearly defined. This is partly due to the energy integration which tends to broaden the observed features, but also to the nature of the spectrum itself as discussed below. The excitations are broad both in ${\bf q}$-space and in energy (see Figure \ref{fig_INS_calc}), which tends to smear out the kagome ice features. At the same time, new dispersive branches form. They stem from the positions of the kagome ice pinch points, spread in reciprocal space and finally close up at about 0.12 meV at ${\bf q}=(2,2,0)$ (see Figure \ref{fig_INS_calc}(c), Supplementary Note 2 and Supplementary Figures 5-10). This set of excitations (kagome ice mode and new dispersive branches) can thus be associated with two-dimensional dynamics of the kagome spins. The change towards this two-dimensional regime occurs progressively, as can be seen from the smooth evolution of the spectra. 
At 0.75 T, a high energy flat mode at about 0.3 meV stands out from these low-energy excitations. This mode can be attributed to the local excitations of the apical spins, which are strongly polarized by the applied field and are not involved in the two-dimensional dynamics. The energy of this mode is thus related to the Zeeman splitting associated to the apical spins. 
 
These observations are consistent with diffraction results shown in Figure \ref{fig_struct}(c). They reveal that the appearance of the kagome ice pattern on the 70 $\mu$eV flat mode, of the new dispersive branches and of the high energy branch matches with the full polarization of the apex spins at $\mu_0 H\ge0.25$~T. Beyond this value, the dimensional reduction driven by the magnetic field confines the fluctuations to the kagome planes, thus transforming the dynamic spin ice mode into a dynamic kagome ice mode. 
 
\section{Discussion}

Theoretically, owing to the dipolar octupolar nature of the \nd\ electronic ground state, the physics of \ndzr\ can be described by an XYZ Hamiltonian \cite{Huang14} written in the local frame of the effective pseudo-spins 1/2, $\boldsymbol{\tau}=(\tau^x,\tau^y,\tau^z)$, residing on each site of the pyrochlore lattice: 
\begin{equation}
\begin{aligned}
{\cal H} =\sum_{<i,j>} &\left[ {\sf J}_x\tau^x_i \tau^x_j + {\sf J}_y \tau^y_i \tau^y_j + {\sf J}_z \tau^z_i \tau^z_j \right.\\
&\left. + {\sf J}_{xz} \left(\tau^x_i \tau^z_j + \tau^z_i \tau^x_j \right) \right] + g_z\mu_{\rm B} \sum_i~{\bf H}.{\bf z}_i~\tau^z_i
\end{aligned}
\label{H}
\end{equation}
${\sf J}_{x}$, ${\sf J}_{y}$, ${\sf J}_{z}$ and ${\sf J}_{xz}$ are effective interactions and $g_z=4.55$ is the effective $g$-factor, deduced from the crystal electric field scheme\cite{Lhotel15} ($g_x=g_y=0$). $\tau^z$ (along $\langle 111 \rangle$) identifies with the dipolar magnetic moment $S^z= g_z \tau^z$. $\tau^{x,y}$ components are non observable quantities which respectively transform as dipolar and octupolar moments under symmetries. 

The main difficulty in describing the zero-field ground state of \ndzr, is to understand the coexistence of an all in -- all out ground state characterized by a reduced moment, with a dynamic spin ice mode. Recently, a plausible scenario has been proposed, assuming that the pseudo-spins ${\bf \tau}$ order in a direction tilted away from their local $z$ magnetic direction, within the $(x, z)$ plane \cite{Benton16b}. This state projects onto the $z$ axes as a classical all in -- all out configuration, but with a reduced moment. The obtained spin excitation spectrum encompasses a dynamic spin ice mode along with dispersive branches, as observed in the experiment. We have studied the evolution of such a state when a magnetic field is applied along $[111]$. We find that the apical spins are polarized and the model predicts the appearance of the kagome ice pattern (see Figure \ref{fig_INS}(b)), as well as of the new dispersion stemming from the kagome ice pinch points (see Figure \ref{fig_INS_calc}). We have analyzed the spectra measured as a function of field and we have found that, to get the best agreement between our data and the model, the set of exchange parameters given in Ref. \citenum{Benton16b} has to be slightly modified. 
This new analysis gives a revised set of parameters: ${\sf J}_x= (-0.36 \pm 0.16)$~K, ${\sf J}_y=(0.066 \pm 0.2)$~K, ${\sf J}_z=(0.86 \pm 0.15)$~K and ${\sf J}_{xz}= (0.44 \pm 0.15)$~K (see Supplementary Notes 3 and 4 and Supplementary Figures 11-14).

As pointed out in Ref. \citenum{Benton16b}, the spin excitation spectrum can be understood by considering the fluctuations of the field emerging from the dynamic components, thus generalizing to the dynamics the concept of emergent field introduced in spin ice \cite{Castelnovo08} (see Supplementary Note 5 and Supplementary Figures 15 and 16). Applying a Helmholtz-Hodge decomposition to these fields gives rise to divergence free and divergence full dynamic fragments, which can be seen as a quantum analog of the magnetic moment fragmentation \cite{Brooks14, Benton16b}. The divergence free part identifies with the flat mode. It is spin ice like in zero field and kagome ice like above 0.25~T. The divergence full part lies in the dispersive branches emerging from the pinch points and corresponds to the propagation of charged quasi-particles. Above 0.25 T, the dimensional reduction confines them to the kagome planes.

The observation of the two-dimensional kagome ice mode over a large field range ($0.25-1$~T) demonstrates the robustness of this feature. Interestingly, field induced transitions in the magnetic structure do not directly affect this low-energy inelastic flat mode, showing that the actual magnetic ordered ground state is well protected from these excitations. In the mean field model presented above, this counterintuitive disconnection is due to the fact that pseudo-spins aligned along the $z$ axes do not give rise to transverse spin excitations visible in neutron scattering. In other words, the observable spectrum originates from fluctuations out of the $(x,y)$ pseudo-spin ordered components, having a non-zero projection onto the magnetic $z$ axes. Remarkably, the resulting flat mode remains at the same energy of about 70 $\mu$eV, in the whole field range from zero to high field, even though its structure factor, and so the nature of the fluctuations are impacted by the field.

While this XYZ model allows us to describe the main features of our observations, it fails in several aspects, which may call for more sophisticated approaches. First, it cannot account for the field induced transition at 0.08~T, possibly due to the high value obtained for the ${\sf J}_{x}$ parameter, which constrains the pseudo-spins along the local $x$ axis. Second, when the field is increased, the energy of the high-energy flat mode corresponding to flipping of the apical spin is shifted in the model towards higher energy than what is observed (see Figure~\ref{fig_INS_calc}). Finally, it predicts well-defined pinch points and excitations, while the experimental features appear much broader than the experimental resolution ($<20~\mu$eV). This is especially true for $H=0$ in the vicinity of ${\bf q}=(-2,1,1)$ and for $H \ne 0$ in the whole ${\bf q}$ range, where the spectrum resembles a continuum.
The broadness of these features is reminiscent of observations in some other pyrochlore frustrated magnets, where the spin wave excitations are also not well defined. A prominent example is Yb$_2$Ti$_2$O$_7$, in which broad spin waves (in some {\bf q} regions) emerge from a continuum when a magnetic field is applied, becoming well defined at large field only \cite{Thompson17}. These results draw attention to the limits of the conventional spin wave approach in frustrated systems in the presence of quantum effects, which are prone to exotic excitations mediated by complex processes. 
In the particular case of \ndzr\, the broadness of the spin waves is likely due to interactions between the divergence free and full dynamical fragments. While decoupled at the mean field level of a spin wave approach, we anticipate that they become coupled in more elaborate treatments. 

Finally, recent theoretical studies of the XYZ Hamiltonian in the presence of a $[111]$ magnetic field discussed above have proposed the existence of a quantum kagome ice phase \cite{Carrasquilla15, Owerre16}. The specific exchange parameters obtained for \ndzr\ locate it quite far from this phase, but the observation of a kagome ice mode in the excitation spectrum paves the way for further explorations. We anticipate that rich physics, from both a theoretical and an experimental point of view, has still to be discovered in this system.


\section*{Methods}

\noindent {\bf Synthesis.} Single crystals of \ndzr\, were grown by the floating-zone technique using a four-mirror xenon arc lamp optical image furnace \cite{monica1,Ciomaga15}. \\
{\bf Inelastic neutron scattering.} Inelastic neutron scattering experiments were carried out at the Institute Laue Langevin (ILL, France) on the IN5 disk chopper time of flight spectrometer and operated with $\lambda~=~8.5$~\AA. The \ndzr\, single crystal sample was attached to the cold finger of a dilution insert and the field was applied along $[111]$. The sample was misaligned by about 2 degrees around the $(1, -1, 0)$ axis and no misalignment could be detected around the $(-1,-1,2)$ axis. The data were processed with the Horace software, transforming the recorded time of flight, sample rotation and scattering angle into energy transfer and ${\bf q}$-wave-vectors. Maps were symmetrized to account for the 6-fold symmetry in this scattering plane. It is worth noting that the magnetic structure factor is not favourable in the scattering plane of these experiments, compared to the case of scattering plane perpendicular to the $[1 \bar{1} 0]$ direction.\\
{\bf Neutron diffraction.} The neutron diffraction data were taken at the D23 single crystal diffractometer (CEA-CRG, ILL France) using a copper monochromator and $\lambda=1.28$ \AA. Here the field was applied along the $[111]$ direction.
Refinements were carried out with the Fullprof software suite \cite{fullprof} on data collections of 60 Bragg peaks. The sample was first saturated in $-3$ T, and data collections were made by increasing the field step by step towards the measurement value. In addition, measurements were performed by collecting the intensity on the top of some Bragg peaks when sweeping the magnetic field from $-1$ to 1 T at a rate of 14.8 mT.min$^{-1}$. \\
{\bf Magnetization.} Magnetization measurements were performed on a SQUID magnetometer equipped with a dilution refrigerator \cite{Paulsen01} on a parallelepiped sample \cite{Lhotel15}. The field was swept from $-0.4$ to 0.4 T. The magnetization is corrected for demagnetization effects. \\
{\bf Calculations.} Calculations were carried out on the basis of a mean field treatment of the XYZ Hamiltonian written in terms of a pseudo-spin $1/2$ spanning the CEF doublet ground state. The spin dynamics were then calculated numerically in the Random Phase Approximation \cite{jensen,kao03,petit14,Robert15}.\\
{\bf Data availability.}  All relevant data are available from the authors. Inelastic neutron scattering data performed at the ILL are available at the doi: 10.5291/ILL-DATA.4-05-637.

\section*{Acknowledgements}
The authors acknowledge P.C.W. Holdsworth for useful discussions and C. Paulsen for the use of his dilution SQUID magnetometer. E. L. acknowledges financial support from ANR, France, Grant No. ANR-15-CE30-0004. The work at the University of Warwick was supported by the Engineering and Physical Sciences Research Council (EPSRC), United Kingdom, through Grant No. EP/M028771/1.

\section*{Author contributions}
Crystal growth and characterisation were performed by MCH, MRL and GB. Inelastic neutron scattering experiments were carried out by SP, EL, JO and HM. Diffraction experiments were carried out by SP, EL and ER. Magnetization measurements were performed by EL. The data were analysed by SP, EL with input from ER and JO. RPA calculations were carried out by SP. The paper was written by EL and SP with feedback from all authors.



\newpage

\newpage

\newcolumntype{M}[1]{>{\centering\arraybackslash}m{#1}}
\newcolumntype{Y}{>{\centering\arraybackslash}X}
\def\tabularxcolumn#1{m{#1}}

\renewcommand{\theequation}{S\arabic{equation}}
\renewcommand{\citenumfont}[1]{S#1}

\makeatletter
\renewcommand{\@biblabel}[1]{\quad S#1. }
\makeatother

\renewcommand{\figurename}{{\bf Supplementary Figure}}
\renewcommand{\tablename}{{\bf Supplementary Table}}
\renewcommand{\thetable}{{\arabic{table}}}
\renewcommand{\thesubsection}{Supplementary Note \arabic{subsection}}
\def\bibsection{\subsection*{\refname}}
\renewcommand{\refname}{Supplementary References}


 \setcounter{figure}{0} 
 \setcounter{equation}{0}

\onecolumngrid
\begin{center} 
{\bf \large{Evidence for dynamic kagome ice - 
Supplementary Information}} 
\end{center}

Additional information about the neutron experiments and their analysis is provided. We explain the analysis of the diffraction experiments, the spin wave model, as well as the fitting procedure used to determine the exchange constants. Finally, we focus on the peculiar properties of the spin wave spectrum, namely its decomposition into two subsets of excitations, that can be described by divergence less and divergence full dynamical emergent fields. 

\vspace{0.4cm}

\begin{table}[h]
\setlength{\extrarowheight}{1pt}
\begin{tabularx}{8.5cm}{M{1.75cm}M{1.75cm}YM{1.5cm}Y} 
\hline
\hline
Site & 1 & 2 & 3 & 4 \\
CEF axis ${\bf z_i}$ & $(1,1,\bar{1})$ & $(\bar{1},\bar{1},\bar{1})$ & $(\bar{1},1,1)$ & $(1,\bar{1},1)$ \\
Coordinates & {\text{\footnotesize$(1/4,1/4,1/2)$}} & {\text{\footnotesize$(0,0,1/2)$}} & {\text{\footnotesize$(0,1/4,3/4)$}} & {\text{\footnotesize$(1/4,0,3/4)$}}\\
\hline 
${\bf a_i}$ & $(\bar{1},\bar{1},\bar{2})$ & $(1, 1,\bar{2})$ & $(1,\bar{1},2)$ & $(\bar{1},1,2)$ \\
${\bf b_i}$ & $(\bar{1},1,0)$ & $(1, \bar{1},0)$ & $(1,1,0)$ & $(\bar{1},\bar{1},0)$ \\
\hline
\hline
\end{tabularx}
\caption{\label{table1} {\bf Site coordinates}. They are written in the cubic $Fd\bar{3}m$ structure of the pyrochlore lattice, for the site dependent local ${\bf a_i}$ and ${\bf b_i}$ vectors spanning the local $({\bf x},{\bf y})$ anisotropy planes. The CEF axes ${\bf z_i}$ of the rare earth ions are perpendicular to these planes.}
\end{table} 

\vspace{-0.2cm}
\begin{figure}[h!]
\includegraphics[height=5cm]{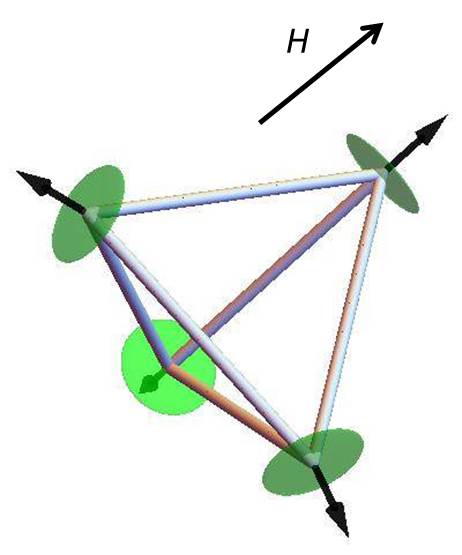}
\caption{\label{Fig1a} {\bf Sketch of a tetrahedron in the pyrochlore lattice. }The black arrows show the local ${\bf z}_i$ anisotropy (crystal field - CEF) axes and the green disks represent the local $({\bf a}_i,{\bf b}_i)$ planes. Further details are given in Supplementary Table~\ref{table1}. }
\end{figure}

\vspace{-1cm}
\begin{figure}[h!]
\includegraphics[height=6.5cm]{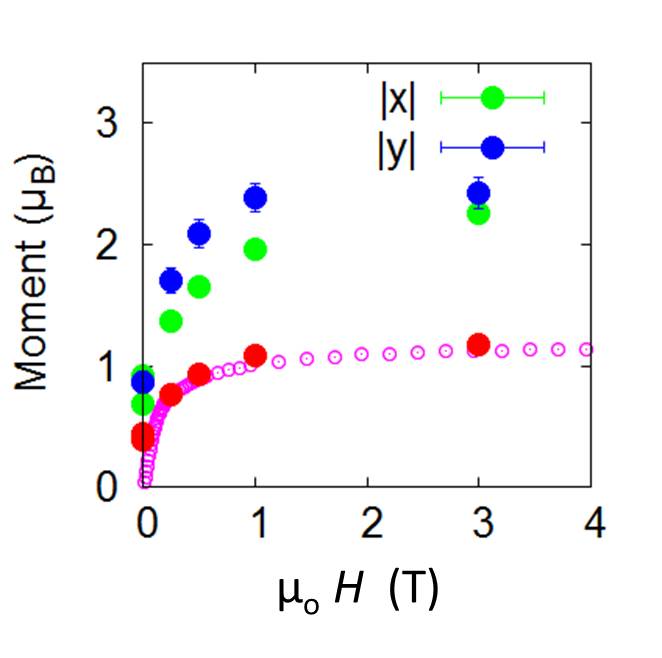}
\caption{\label{Fig1b} {\bf Field dependence of the magnetic moments in the unit cell for ${\bf H} \parallel [111]$. }$y$ (in blue) gives the moment of the apical spin, with its CEF axis along the field. $x$ (in green) gives the moment of the three remaining spins (also called kagome). The magenta curve displays the magnetization obtained from macroscopic measurements. Red points correspond to the magnetization calculated using the refinements results and projected onto the field direction, leading to $M(H)=(x-y)/4$. The error bars are provided by the Rietveld refinement made using Fullprof.}
\end{figure}

\clearpage


\subsection{Neutron diffraction}

{\bf Experimental details: } The neutron diffraction data were collected using the D23 single crystal diffractometer (CEA-CRG, ILL France) operated with a copper monochromator and using $\lambda=1.28$ \AA. The single crystal of \ndzr\ was glued on the Cu finger of a dilution insert and placed in a cryomagnet. The experiments have been conducted with the (vertical) field ${\bf H}$ parallel to the $[111]$ direction. 
We first checked that the Bragg intensities remain zero on the forbidden ${\bf q}$ vectors of the F centred lattice. This implies that the field induced structures are described by a ${\bf k} = {\bf 0}$ propagation vector. The additional magnetic intensity is then observed on top of the nuclear Bragg peaks. A series of integrated intensities at 10 K and $H=0$ was measured (about 60 reflections) as a reference. Refinements were performed using the {\sc Fullprof} sofware suite \cite{fullprof}, working on the differences with the $T=10$~K data set. 
Finally, for a set of chosen Bragg peaks, we ramped the field back and forth, up to 1 and back to -1 T, yielding a precise evolution of the magnetic intensities. The 1 T field was chosen to saturate the sample.\\

{\bf Model of the magnetic structure in a $[111]$ field: }
Since the propagation vector is ${\bf k} = {\bf 0}$, the four tetrahedra of the crystalline unit cell host identical arrangements of the spins. Furthermore, because of the strong Ising anisotropy of the \nd\ ions, it is reasonable to consider that the magnetic moments are forced to align along the local CEF axis (labelled ${\bf z}_i$, see Supplementary Figure \ref{Fig1a} and Table \ref{table1}). 
In our refinement, we thus assume the following model (see Table \ref{table1}):
\begin{equation}
\begin{array}{lcl}
{\bf m}_2  &=& y~{\bf z}_2 \\
{\bf m}_{i} &=& x~{\bf z}_{i}, ~~~i=1,3,4
\end{array}
\end{equation}
where $x$ and $y$ are the fitted parameters. Note that ${\bf m}_2$ is parallel to $[111]$, hence to the applied field. From a physical point of view, situations where $x$ and $y$ have the same sign correspond to a generalized all in -- all out structure. When $x$ and $y$ have different signs, however, the model describes a structure which resembles the 1 out -- 3 in, but ${\bf m}_2$ and ${\bf m}_{1,3,4}$ may have different amplitudes. The magnetization per \nd\ ion is given by:
\begin{equation}
{\bf M} = \frac{1}{4} \sum_{i \in \boxtimes =\left\{1,2,3,4\right\}} {\bf m}_i = \frac{(x-y)}{4\sqrt{3}}\left(
\begin{array}{c}
1\\
1\\
1
\end{array}
\right)
\end{equation}
Projected along the $[111]$ field, it simply becomes $M=(x-y)/4$. The results obtained increasing the field from 0 to 3 T along with the calculated magnetization are shown in Supplementary Figure \ref{Fig1b}. The excellent agreement with the macroscopic magnetization is also shown for comparison. It is worth noting that the magnetic structure may also be described in terms of generalized charges ${\bf Q}_{\boxtimes}$ living on the dual (diamond) lattice defined by the centres of the tetrahedra. ${\bf Q}_{\boxtimes}$ is a vector with three components defined by:
\begin{equation}
\begin{aligned}
Q^x_{\boxtimes}=&\sum_{i \in \boxtimes=\left\{1,2,3,4\right\}} {\bf m}_i.{\bf a}_i\\
Q^y_{\boxtimes}=&\sum_{i \in \boxtimes=\left\{1,2,3,4\right\}} {\bf m}_i.{\bf b}_i\\
Q^z_{\boxtimes}=&\sum_{i \in \boxtimes=\left\{1,2,3,4\right\}} {\bf m}_i.{\bf z}_i = 3x+y
\end{aligned}
\end{equation}
In zero field, the all in -- all out structure can be considered as a $Q^z$ charged staggered pattern. \\

{\bf Analysis of the ramps}:
We now turn to the analysis of the intensities measured while ramping the magnetic field (see Supplementary Figure \ref{Fig-QH}). To this end, we first write the magnetic structure factor for a wavevector ${\bf q}$ :
\begin{equation}
{\bf F}_{\rm M}({\bf q}) = \sum_i {\bf m}_i~f_i~e^{i {\bf q} {\bf R}_i}
\end{equation}
where ${\bf R}_i$ denotes the position of the $i^{\rm th}$ spin in the unit cell and $f_i$ is the form factor. The magnetic Bragg intensity is then given by: $I_{\bf q} = \sum_{a,b=x,y,z} {\bf F}^a_{\rm M}({\bf q}) \left(\delta_{a,b}-\frac{q^a q^b}{q^2}\right) {\bf F}^b_{\rm M}({\bf q})^*$. For ${\bf q}=(2,0,0)$, $(1,\bar{1},1)$ and $(1,3,\bar{1})$, analytic calculations show that:
\begin{equation}
\begin{array}{lcl}
I_{200}-I_{200}^{\rm N}     & = & 32 V (x-y)^2 \\
I_{1\bar{1}1}-I_{1\bar{1}1}^{\rm N} & = & 42.66 V (x-y)^2 \\
I_{13\bar{1}}-I_{13\bar{1}}^{\rm N} & = & V\left(A x^2 + 2B x y + C y^2\right)
\end{array}
\end{equation}
where $V$ is the volume, $I^{\rm N}_{\bf q}$ denote the corresponding nuclear contributions and the coefficients are given by: $A = 174.55$, $B = -11.64$ and $C=34.91$. In these expressions, the form factor has been neglected since it is close to 1 for those ${\bf q}$ vectors. It is then convenient to consider the new intermediate variables $e$ and $t$ defined as:
\begin{equation}
\label{variables}
\begin{aligned}
e &=&y-x \\
t& =&y/x-1\\
&\Rightarrow& e=tx
\end{aligned}
\end{equation}
hence:
\begin{equation}
\begin{aligned}
I_{200}-I_{200}^{\rm N}     & =  32 V e^2 \\
I_{1\bar{1}1}-I_{1\bar{1}1}^{\rm N} & =  42.66 V e^2 \\
I_{13\bar{1}}-I_{13\bar{1}}^{\rm N} & =  V \left( (A+ 2 B + C) + 2(B+C) t + C t^2 \right) \frac{e^2}{t^2} \qquad {\rm when} \: t,e\neq0. 
\end{aligned}
\end{equation}
Note that $e=0$ at 0 T since the structure is all in -- all out ($x=y$). We now consider reduced quantities where the $o$ and $1$ superscripts denote respectively the values at 0 and 1 T. We then have $I_{1\bar{1}1}^{\rm o} = I_{1\bar{1}1}^{\rm N}$ and $I_{200}^{\rm o}=I_{200}^{\rm N} $. By introducing $\zeta = I_{13\bar{1}}^{\rm o}-I_{13\bar{1}}^{\rm N} = V\left(A +2B+C\right) x_o^2$, $\mu=1/e_1^2$ and $\eta=I_{13\bar{1}}^1-I_{13\bar{1}}^{\rm o}$,
\begin{equation}
\begin{aligned}
i_{200} & =  \frac{I_{200}-I_{200}^{\rm o}}{I_{200}^1-I_{200}^{\rm o}} = \frac{e^2}{e_1^2}= \mu e^2\\
i_{1\bar{1}1} & =  \frac{I_{1\bar{1}1}-I_{1\bar{1}1}^{\rm o}}{I_{1\bar{1}1}^1-I_{1\bar{1}1}^{\rm o}}= \frac{e^2}{e_1^2}= \mu e^2\\
i_{13\bar{1}} & =  \frac{I_{13\bar{1}}-I_{13\bar{1}}^{\rm o}}{I_{13\bar{1}}^1-I_{13\bar{1}}^{\rm o}} \\
 & = \eta \left[ \frac{e^2}{t^2}\left((A+ 2 B + C) + 2(B+C)t + C t^2\right) - \zeta \right].
 \end{aligned}
\end{equation}
As a result, the variable $e$ is readily obtained from the field evolution of the $(2,0,0)$ or $(1,\bar{1},1)$ intensity:
\begin{equation}
e^2 = i_{200}/\mu.
\end{equation}
Note that the data are consistent, i.e. $i_{200}$ and $i_{1\bar{1}1}$ give the same information. Finally, we need to solve the following equation, deduced from the expression for $i_{13\bar{1}}$:
\begin{equation}
0 = t^2 \left(1-\frac{C e^2}{i_{13\bar{1}}/\eta+\zeta}\right) - 2 t (B+C)
\frac{e^2}{i_{13\bar{1}}/\eta+\zeta} - (A+2B+C)\frac{e^2}{i_{13\bar{1}}/\eta+\zeta}.
\end{equation}
For each field value, or in other words, for each $e$ value implicitly determined from $i_{200}$, there exist two solutions $t_1$ and $t_2$, hence two solutions for the $y/x=(1+t)$ ratio. These solutions take the form of the black and grey branches shown in Supplementary Figure \ref{Figure2}a. The final choice between the two possibilities relies on physical grounds and is shown by the red points in Supplementary Figure \ref{Figure2}a: when the field is strongly negative or positive, a 3 in -- 1 out structure is expected, i.e. $y/x \le 0$. This corresponds to the black branch in Supplementary Figure \ref{Figure2}a. In zero field, the all in -- all out structure is stabilized, which corresponds to $y/x =1$, hence to the point where the black and grey solutions coincide in Supplementary Figure \ref{Figure2}b. Since $x$ and $y$ should be regular functions of the field, except close to the transition at 0.08 T where a discontinuity can be expected, the physical solution stays on the black branch. This holds up to the transition at 0.08 T. Above, there is a jump to join the grey solution. This gives the field evolution for the $x$ and $y$ parameters shown in Supplementary Figure \ref{Figure2}c. The blue arrow marks the discontinuity at the transition.\\


\begin{figure*}[h!]
\includegraphics[width=\textwidth]{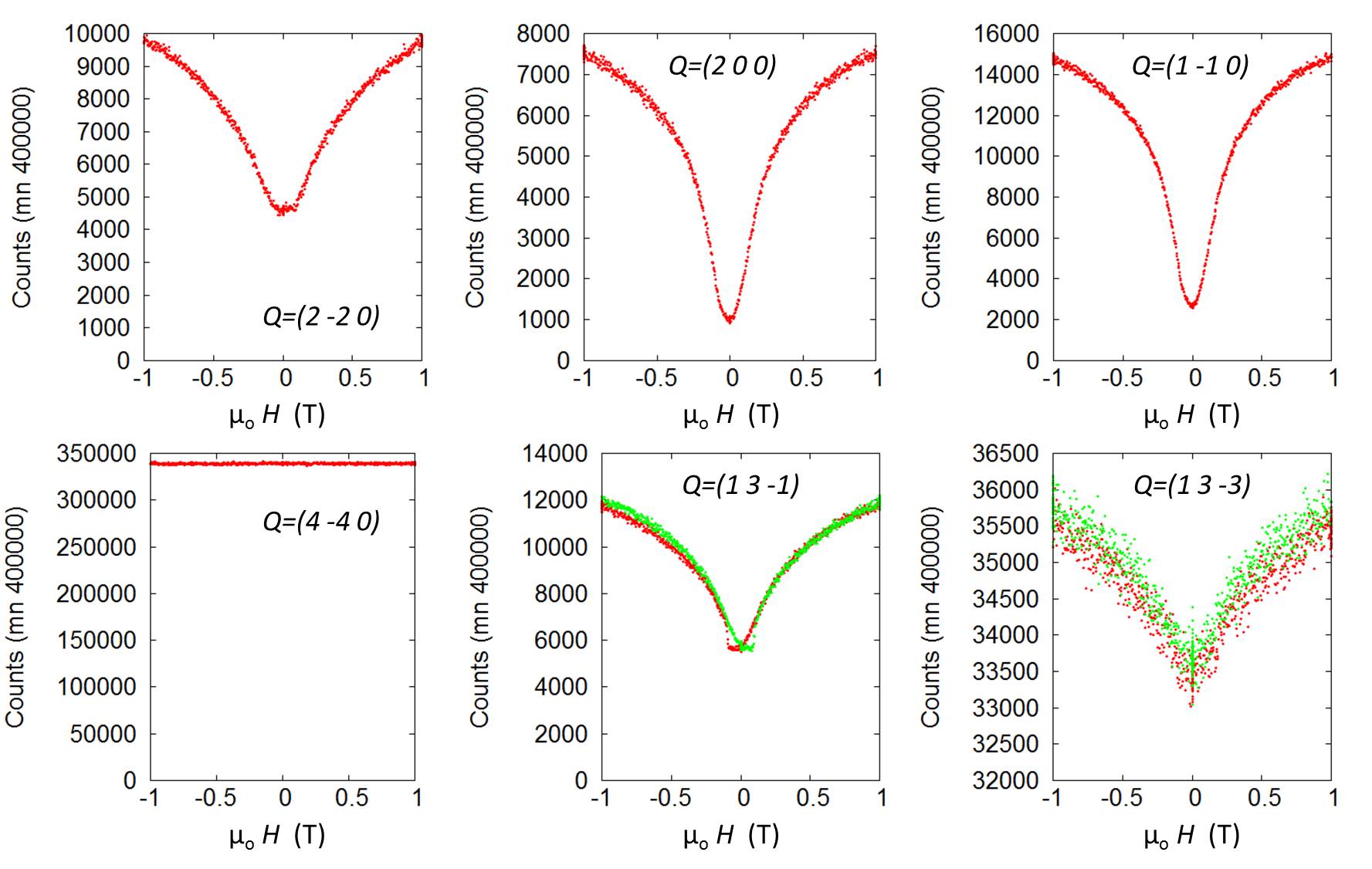}
\caption{{\bf Bragg peak intensities vs $H$, with the field applied along $[111]$}. They were measured while ramping the field back and forth up to 1 and back to $-1$ T. Red and green points correspond to sweeping the field from positive and negative values respectively.}
\label{Fig-QH}
\end{figure*}


\begin{figure*}[h!]
\includegraphics[width=\textwidth]{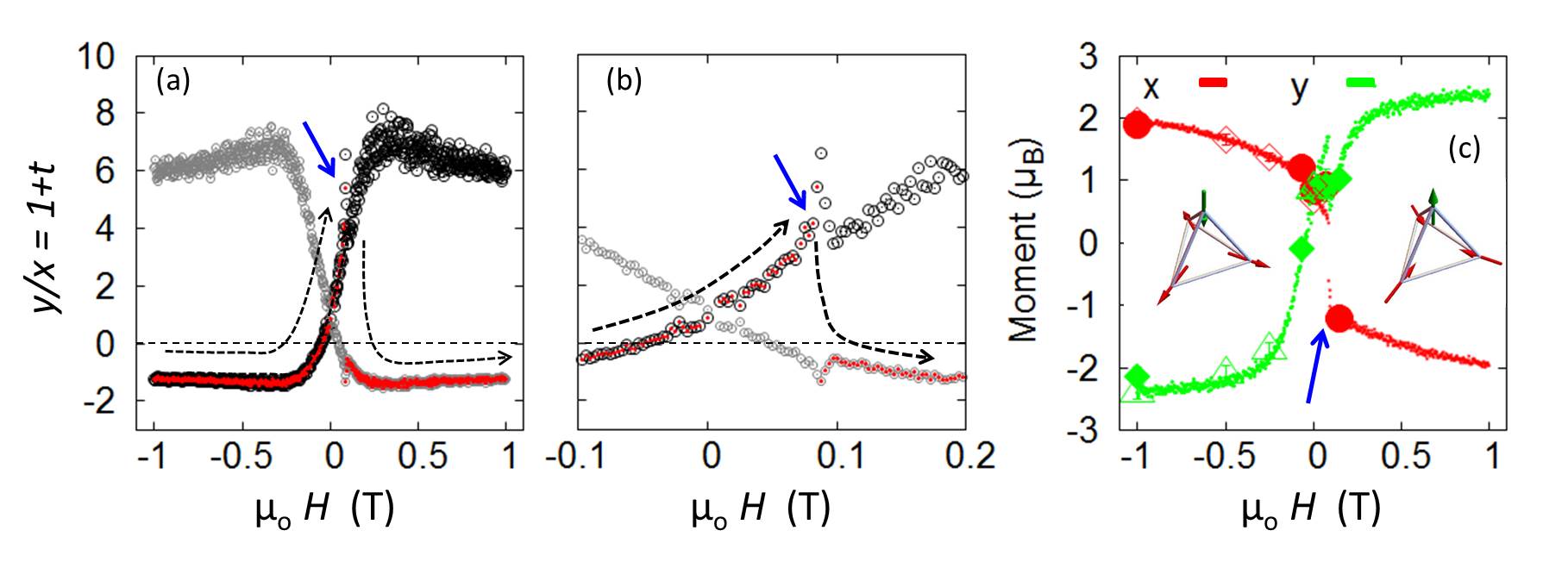}
\caption{{\bf Field dependence of the variables defined in Supplementary equation \ref{variables}.} Figures (a) and (b) show $y/x=(1+t)$ as a function of the field. Two branches (grey and black) are found. The blue arrow marks the transition and the dotted black arrows the choice between the two branches. (c) shows the field dependence of $x$ and $y$ (closed symbols). As a complement, data were taken with a graphite PG monochromator, reported here with open symbols. Note the satisfying agreement with the Fullprof refinement.}
\label{Figure2}
\end{figure*}

\clearpage

\subsection{Spin dynamics  and inelastic neutron scattering}

Inelastic neutron scattering experiments were carried out on the IN5 disk chopper time of flight spectrometer operated by the Institute Laue Langevin (ILL, France). The field was applied along 
$[111]$ and the \ndzr\, single crystal mounted to have the $(-k,k,0)$-$(-h,-h,2h)$ reciprocal directions in the horizontal scattering plane. A sketch of the $(-k,k,0)$-$(-h,-h,2h)$ scattering plane along with the positions of remarkable wavevectors is displayed in Supplementary Figure \ref{sqw0}. The kagome ice pinch points (in yellow) and all in -- all out positions (in blue) are especially shown. \\

As a very good energy resolution is needed (about 20 $\mu$eV), we used a wavelength $\lambda=8.5$ \AA. The data were then processed with the {\sc horace} software \cite{horace}, transforming the recorded time of flight, sample rotation and scattering angle into energy transfer and $q$-wave-vectors. The offset of the sample rotation was determined based on the Bragg peak positions. In all the experiments, the sample was rotated in steps of 1 degree. We finally prepared a set of constant ${\bf q}$ scans by integrating over a small ${\bf q}$ range with $\Delta h=\Delta k=0.05$. We also prepared a series of constant energy maps (ranging from $E=0.05$ up to 0.25 meV) that were symmetrized on the basis of the expected 6-fold symmetry of the intensity in this scattering plane. These maps are displayed in Supplementary Figure \ref{sqw1} and \ref{sqw2}, giving an overview of the spin dynamics for a series of applied magnetic fields. Note that those maps were obtained by integrating over a range $\Delta \omega = 10~\mu$eV, which is about half the energy resolution. Supplementary Figure \ref{sqw3} shows the corresponding ${\bf q}$-integrated spectra. Finally, Supplementary Figure \ref{sqw4} and \ref{sqw5} display the dispersions measured along the black dotted and full slices shown in Supplementary Figure \ref{sqw0}.\\

Supplementary Figures \ref{sqw3} to \ref{sqw5} show that there is little evolution of the scattering at low field, i.e. for $\mu_0H=0,~0.07$ and $0.15$~T. A dispersive branch (marked by the A symbol) is identified in the Supplementary Figures \ref{sqw4} and \ref{sqw5}. While the top of this branch clearly appears at $(-2,2,0)$ and symmetry related points (with an energy of 0.25 meV), it is difficult to determine from these dispersions, whether or not this branch reaches a minimum. A close look at Supplementary Figure \ref{sqw1} shows that the branch does reach a minimum at the $(-4/3,2/3,2/3)$-like points, with an energy of about 0.1 meV, thus above the energy of the flat mode. 
%
This can be observed in the two left columns of Supplementary Figures \ref{sqw1} and \ref{sqw2}, and is marked by the A symbol on Supplementary Figures \ref{sqw4} and \ref{sqw5}. \\

Above $\mu_0H=0.25$ T, a different picture sets in. First of all, the structure factor of the flat mode evolves: the intensity close to ${\bf q}=0$ strongly decreases while 6 arms take shape. With increasing field, the intensity of the arms strongly decreases (see the first row of Supplementary Figure \ref{sqw1}; note also the change in the color-scale). In addition, two narrow and dispersive branches, labelled A and B, can be observed (see Supplementary Figures \ref{sqw4} and \ref{sqw5}). The A branch is descended from the zero field dispersive branch. 
For larger fields, this branch is still there but is quite difficult to observe since it becomes extremely weak. However, by taking the ${\bf q}$-average of the scattering (see Supplementary Figure \ref{sqw3}), which basically gives the density of states (enhancing flat modes), it becomes clearly visible. From this data, we can conclude that the A branch bandwidth becomes narrower, and progressively shifts towards higher energies. \\
 
The second dispersing B branch 
emerges from the kagome ice pinch point positions ($(-4/3, 2/3,2/3)$-like points) above the 6 arms mode
and closes at $(-2,2,0)$ (and symmetry related points). With increasing the field, this B branch develops up to about 0.10 meV at $\mu_0H=0.25$ T. The top of the band slightly shifts to higher energies with further increasing the energy. At $\mu_0H=0.75$~T, for instance, it shifts up to 0.13 meV.

\begin{figure*}[h]
\includegraphics[height=7.5cm]{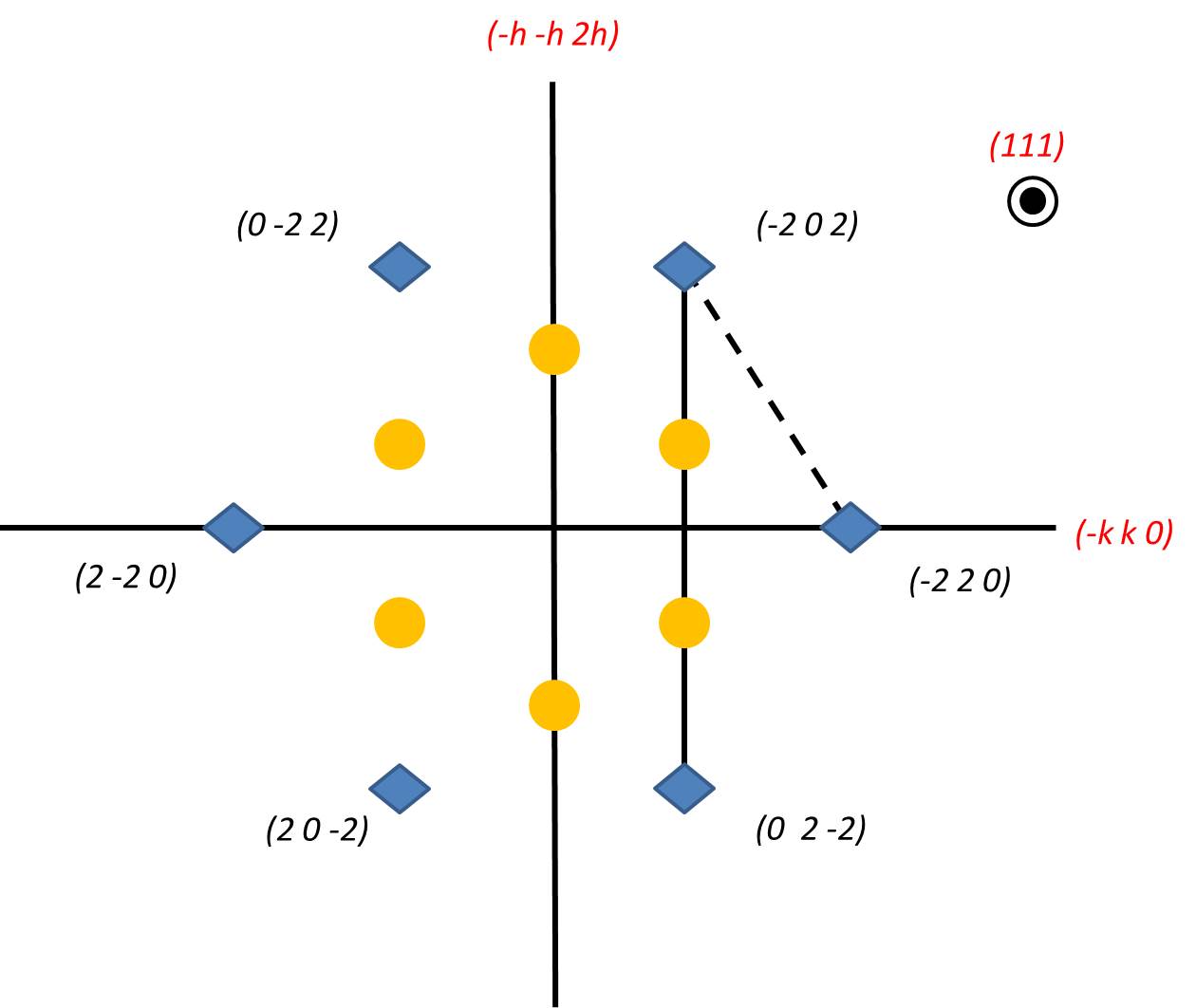}
\caption{{\bf Sketch of the $(-k,k,0)$ $(-h,-h,2h)$ scattering plane for a $[111]$ vertical magnetic field axis}. The blue diamonds correspond to the all in -- all out Bragg peak positions observed in zero field. The yellow points show the positions of the pinch points arising in the kagome ice model. These positions are of the form $(-4/3, 2/3,2/3)$ and symmetry related positions. Black dotted line and the full line show the direction of the maps shown in Supplementary Figures \ref{sqw4} and \ref{sqw5} respectively.}
\label{sqw0}
\end{figure*}


\begin{figure*}[h]
\includegraphics[height=7cm]{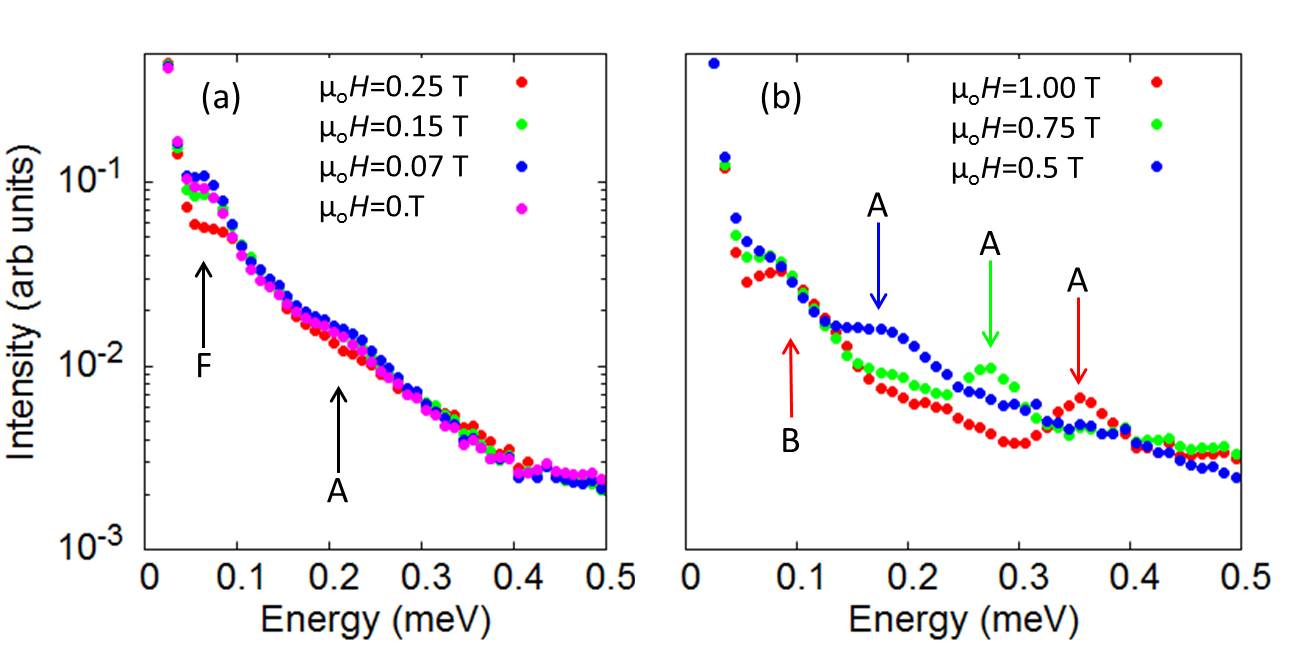}
\caption{{\bf $q$-integrated spectra for a series of magnetic fields applied along $[111]$}.  In (a), the arrows labelled $F$ and $A$ correspond respectively to the flat mode and to the A dispersive mode described in the text. In (b), the coloured arrows show the position of the A and B branches for different fields. Both branches shift to higher energies with increasing field.}
\label{sqw3}
\end{figure*}


\begin{figure*}[h]
\includegraphics[scale=0.35]{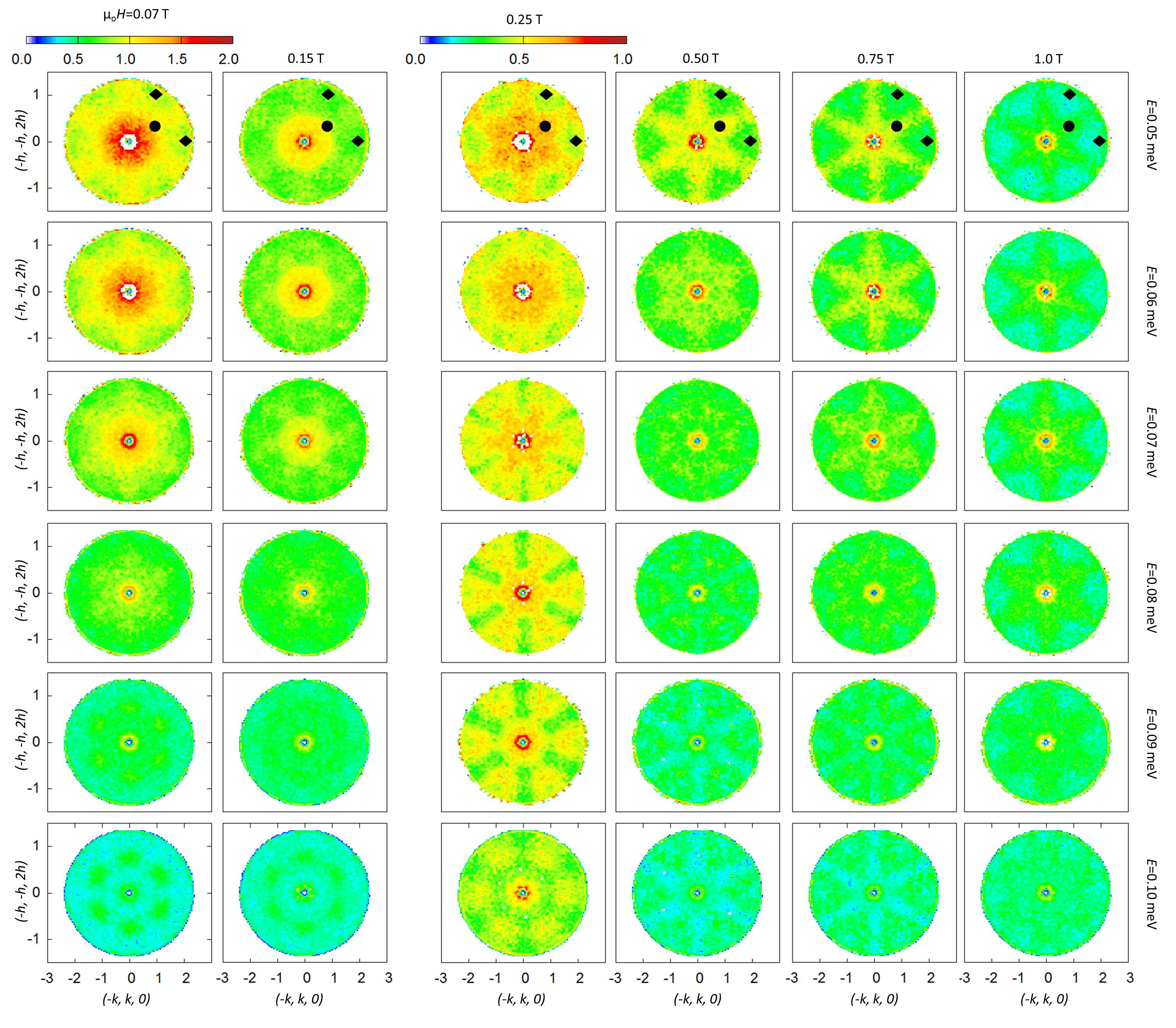}
\caption{{\bf Constant energy maps in the $(-k, k, 0)$-$(-h, -h, 2h)$ scattering plane at energies up to 0.1 meV and for fields between 0 and 1 T.}  They were measured at 60 mK. The black point marks the $(-4/3,2/3,2/3)$ position, and the black diamonds mark the $(-2,2,0)$ and the $(-2,0,2)$ positions.}
\label{sqw1}
\end{figure*}


\begin{figure*}[h]
\includegraphics[scale=0.35]{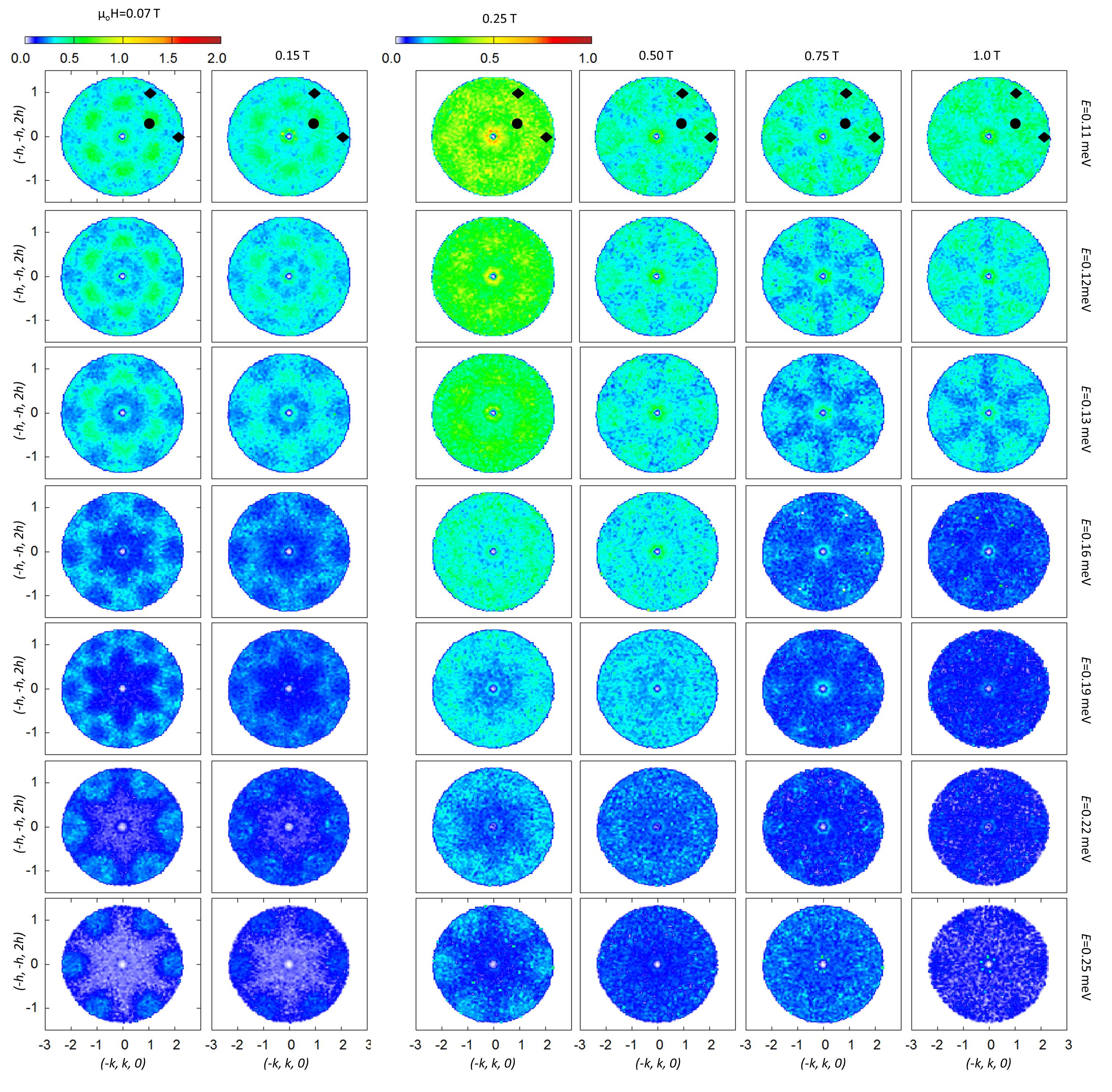}
\caption{{\bf Constant energy maps in the $(-k, k, 0)$-$(-h, -h, 2h)$ scattering plane at energies between 0.11 and 0.25 meV and for fields between 0 and 1 T}. They were measured at 60 mK. The black point marks the $(-4/3,2/3,2/3)$ position, and the black diamonds mark the $(-2,2,0)$ and the $(-2,0,2)$ positions.}
\label{sqw2}
\end{figure*}


\begin{figure*}[h]
\includegraphics[width=\textwidth]{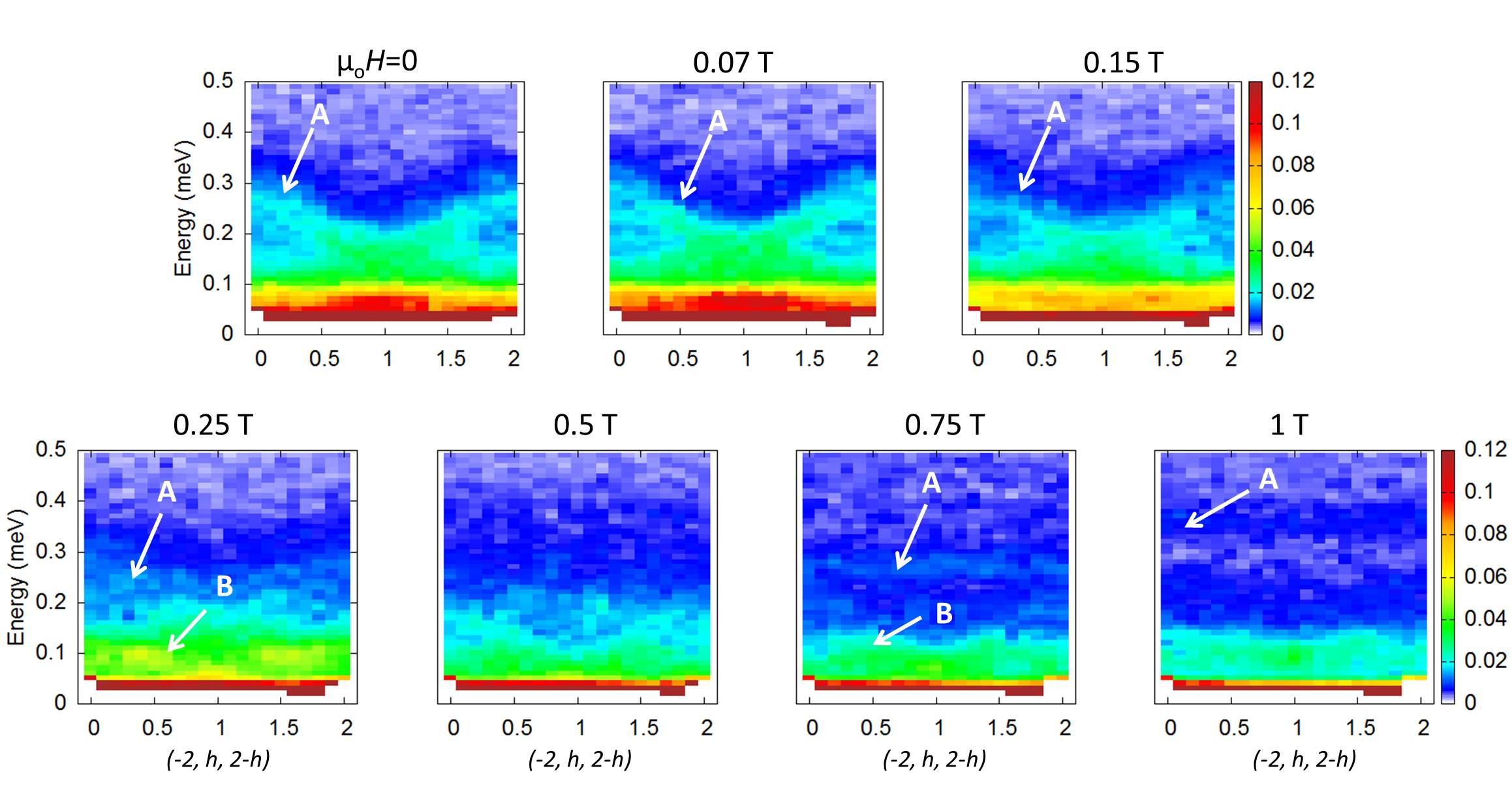}
\caption{{\bf Dispersions along $(-2,h,2-h)$ measured for different fields between 0 and 1 T at 60~mK.} See the black dotted line in Supplementary Figure \ref{sqw0}. $\mu_0H=0.25$ T is the threshold which clearly separates two regimes.}
\label{sqw4}
\end{figure*}


\begin{figure*}[h]
\includegraphics[width=\textwidth]{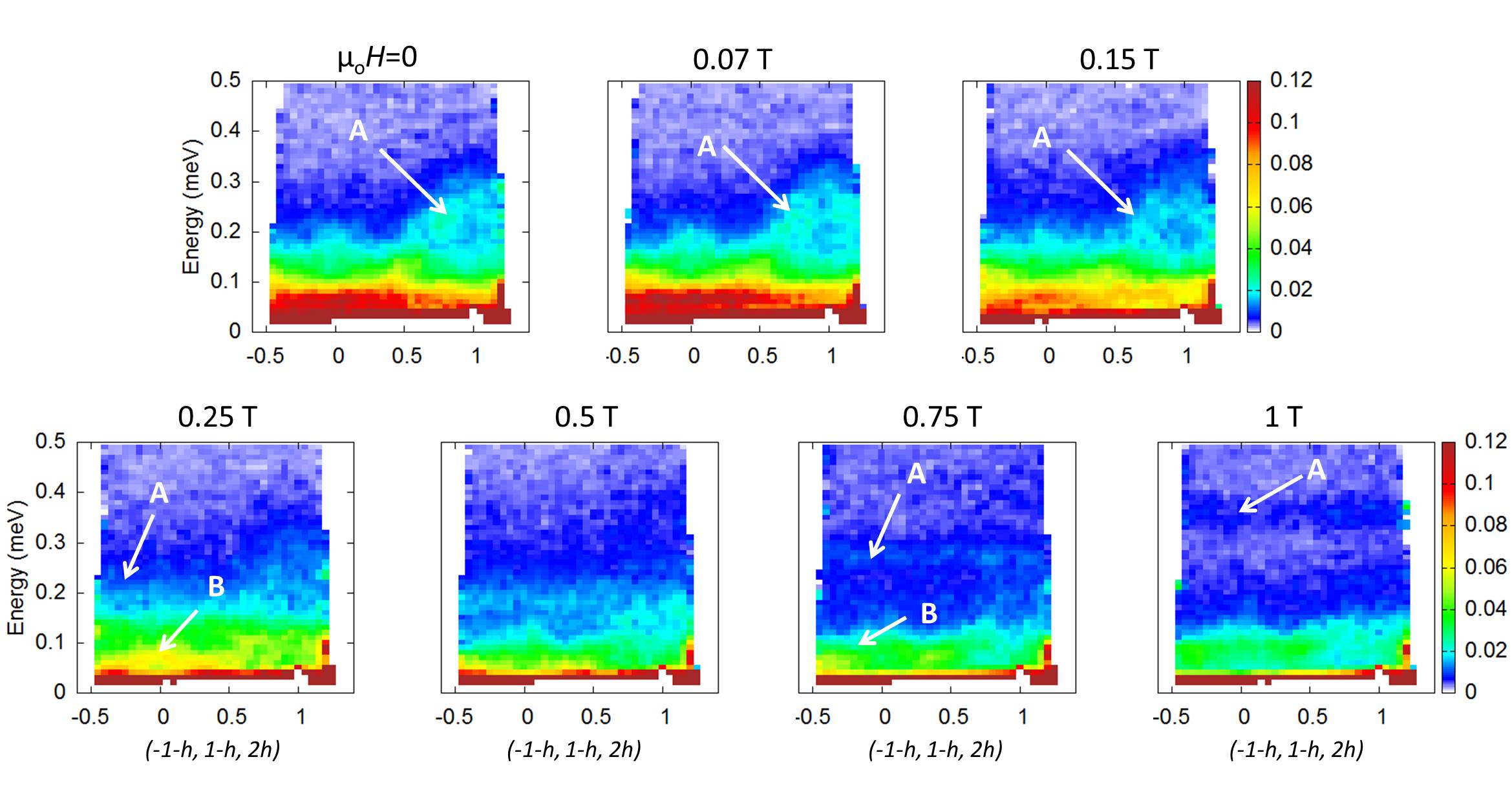}
\caption{{\bf Dispersions along $(-1-h,1-h,2h)$ measured for different fields between 0 and 1~T at 60~mK.} See the black line in Supplementary Figure \ref{sqw0}. }
\label{sqw5}
\end{figure*}
\clearpage

\subsection{The XYZ model}

Following literature \cite{Huang14,Petit16,Benton16b}, the relevant model to describe the case of \ndzr\, is the XYZ Hamiltonian:
\begin{equation}
{\cal H} = \sum_{\langle i,j \rangle} \left({\sf J}_{x} \tau^x_i \tau^x_j + {\sf J}_{y} \tau^y_i \tau^y_j + {\sf J}_{z} \tau^z_i \tau^z_j + {\sf J}_{xz} (\tau^x_i \tau^z_j+\tau^z_i \tau^x_j) \right).
 \label{hxyz}
\end{equation}
In this expression, $\tau_i$ is a pseudo-spin that resides on the sites of the pyrochlore lattice. The $({\bf x}, {\bf y}, {\bf z})$ coordinates refer to the site dependent frames, where ${\bf z}$ is the local $\langle 111 \rangle$ axis (see also Supplementary Table \ref{table1}). Note that the ${\bf z}$ component is related to the physical magnetic moment ${\bf m}=g_{\parallel}~ \tau^z~{\bf z}$, while the ${\bf x}$ and ${\bf y}$ components are non observable quantities. They respectively transform as dipolar and octupolar moments under symmetries. A rotation by an angle $\theta$ in the $({\bf x}, {\bf z})$ plane allows one to define a new $(\tilde{{\bf x}}, \tilde{{\bf z}})$ frame, new pseudo-spin components $(\tilde{\tau}^{\tilde{x}}, \tilde{\tau}^{\tilde{y}}, \tilde{\tau}^{\tilde{z}})$ and new coupling constants $(\tilde{{\sf J}}_{x}, \tilde{{\sf J}}_{y}, \tilde{{\sf J}}_{z})$ so that the Hamiltonian becomes diagonal: 
\begin{equation}
{\cal H} = \sum_{\langle i,j \rangle} \left(\tilde{{\sf J}}_{x} \tilde{\tau}^{\tilde{x}}_i \tilde{\tau}^{\tilde{x}}_j + \tilde{{\sf J}}_{y} \tilde{\tau}^{\tilde{y}}_i \tilde{\tau}^{\tilde{y}}_j + \tilde{{\sf J}}_{z} \tilde{\tau}^{\tilde{z}}_i \tilde{\tau}^{\tilde{z}}_j \right).
\end{equation}
The rotation matrix is defined as:
\begin{equation}
{\cal R} = 
\left(
\begin{array}{ccc}
\cos \theta & & \sin \theta \\
& 1 & \\
-\sin \theta & & \cos \theta
\end{array}
\right)
~\mbox{and}~
\left(
\begin{array}{ccc}
{\sf J}_{x} & & {\sf J}_{xz} \\
& {\sf J}_{y} & \\
{\sf J}_{xz} & & {\sf J}_{z}
\end{array}
\right)
= {\cal R}^T 
\left(
\begin{array}{ccc}
\tilde{{\sf J}}_{x} & & \\
& \tilde{{\sf J}}_{y} & \\
& & \tilde{{\sf J}}_{z}
\end{array}
\right)
{\cal R}
\end{equation}
With this definition, the angle between the ${\bf z}$ and $\tilde{\bf z}$ axes is $\theta$. The following relations are especially useful:
\begin{equation}
\begin{aligned}
\tan 2\theta =& \frac{2 {\sf J}_{xz}}{{\sf J}_{x}-{\sf J}_{z}}, \\
{\sf J}_{xz} =& (\tilde{{\sf J}}_{x}-\tilde{{\sf J}}_{z}) \sin \theta \cos \theta, \\
{\sf J}_{x} =& \tilde{{\sf J}}_{x} \cos^2 \theta + \tilde{{\sf J}}_{z} \sin^2 \theta, \\
{\sf J}_{z} =& \tilde{{\sf J}}_{z} \cos^2 \theta + \tilde{{\sf J}}_{x} \sin^2 \theta.
\end{aligned}
\end{equation}
Importantly, the energies of the spin wave modes solely depend on the values of the three parameters $\tilde{{\sf J}}_{x}, \tilde{{\sf J}}_{y}$ and $\tilde{{\sf J}}_{z}$. This does not mean that ${\sf J}_{xz}$ or the angle $\theta$ do not play any role; the latter especially appears when projecting the pseudo-spin onto the local ${\bf z}$ magnetic direction. As a result, it enters the spin correlation functions (i.e. the inelastic neutron scattering cross section) as well as the magnetization along with its field dependence.\\

In Supplementary Reference \citenum{Petit16}, we assumed ${\sf J}_{x}={\sf J}_{xz}=0$, following arguments based on the analysis of the crystal field. Furthermore, we found that: 
\begin{equation}
{\sf J}_{y} = -0.55~{\rm K,}~ {\sf J}_{z} = 1.2~\mbox{K}
\end{equation} 
allow us to reproduce quantitatively the INS spectra. These parameters also led to an all in -- all out ordering along the $y$ octupolar axis. Note that at the classical level, this particular ordering is stabilized provided that:
\begin{equation}
{\sf J}_{y}~\leq {\sf J}_{z}~\leq~3 |{\sf J}_{y}|~~\mbox{and}~~{\sf J}_{y}~\leq~0.
\end{equation}
Spin wave calculations showed then that the spectrum encompasses two contributions, a flat mode $E_o$ at:
\begin{equation}
E_o = \sqrt{3 |{\sf J}_{y}|(3 |{\sf J}_{y}|-{\sf J}_{z})}
\label{eo1}
\end{equation}
along with dispersive branches, in excellent agreement with INS results. It is worth emphasizing that this analysis, however, does not explain the origin of the {\it magnetic} all in -- all out ordering (along the ${\bf z}$ axis). It was actually designed in the framework of the fragmentation scenario, following the idea that the magnetization can be decomposed into two independent fragments, the divergence full and the divergence free emergent fields of a Helmholtz-Hodge decomposition. The above description, and especially the condensation of the octupolar ordering, was thus thought to apply to the divergence free part only. \\

In Supplementary Reference \citenum{Benton16b}, O. Benton followed a different route. Taking advantage of a global symmetry of the Hamiltonian, he proposed to inter-change the axes compared to Supplementary Reference \citenum{Petit16} so that: 
\begin{equation}
\tilde{{\sf J}}_{x} = 1.2~{\rm K},~
\tilde{{\sf J}}_{y} = 0,~
\tilde{{\sf J}}_{z} = -0.55 ~\mbox{K,}
\end{equation}
and introduced a non zero angle $\theta = 0.83 (48^{\circ})$ chosen to reproduce the Curie-Weiss temperature. This set of parameters corresponds to: 
\begin{equation}
{\sf J}_{x} = 0.247~{\rm K},~
{\sf J}_{y} = 0,~
{\sf J}_{z} = 0.403~{\rm K},~
{\sf J}_{xz} = 0.871~\mbox{K.}
\end{equation}
For this choice, the classical ground state is an all in -- all out configuration with respect to the rotated $\tilde{{\bf z}}$ axis. The spins $\langle \tau \rangle$ thus point along the $\tilde{{\bf z}}$ axes. This rotated all in -- all out order has projections onto the original magnetic ${\bf z}$ axis: 
\begin{equation}
m = g_{\parallel}~\frac{\cos \theta}{2}
\end{equation}
and along the ${\bf x}$ axis: 
\begin{equation} 
\langle \tau_x \rangle \sim 1/2 \sin\theta
\end{equation}
(note again that the latter cannot be observed directly with neutrons). INS results are equally well reproduced with these parameters since the spin wave energies do not depend on $\theta$.\\
 
\subsection{Fitting the INS data}

In the following, we adopt the scheme discussed immediately above and in Supplementary Reference \citenum{Benton16b}, but revisit the values of the parameters to fit the data taken in zero and in applied field. Our initial choice of $\theta$ is different, and is based on the following relation:
\begin{equation}
\cos \theta = \frac{m^{\rm ord}}{m^{\rm sat}} \approx 0.357
\end{equation}
where $m^{\rm ord} = 0.8\mu_{\rm B}$ is the experimental all in -- all out ordered moment and $m^{\rm sat}=2.28$
 is the calculated saturation moment determined from crystal field coefficients \cite{Lhotel15}. This corresponds to an angle:
\begin{equation}
\theta \approx 69^{\circ}.
\end{equation}
To ensure that an all in -- all out state is stabilized along the $\tilde{{\bf z}}$ axis, we assume that the coupling constants belong to the domain ${\cal D}$ defined by 
\begin{equation}
\tilde{{\sf J}}_{z}~\leq~\tilde{{\sf J}}_{x},~\tilde{{\sf J}}_{y} ~\leq 3 |\tilde{{\sf J}}_{z}|~~\mbox{and}~~\tilde{{\sf J}}_{z} ~\leq~0.
\end{equation}
For couplings within ${\cal D}$, the spin wave calculations \cite{Petit16,Benton16b} show that the spectrum encompasses a flat mode $E_o$ at:
\begin{equation}
E_o = \sqrt{(3 |\tilde{{\sf J}}_{z}|-\tilde{{\sf J}}_{x})( 3 |\tilde{{\sf J}}_{z}|-\tilde{{\sf J}}_{y})}
\label{eo}
\end{equation}
along with dispersive branches. To determine the values compatible with our experiments, we let the three parameters vary within ${\cal D}$, for the fixed value of $\theta$ given above. Our goodness criterion is based on a $\chi^2$ defined by the squared difference between the calculated and measured spin wave energies. In our definition, $\chi^2$ takes into account ${\bf q}$ points along $(-2,h,2-h)$ at $H=0$ and 0.75 T. The inverse $\chi^2$ is shown in Supplementary Figure \ref{chi2}. We find that the spin wave spectrum is well reproduced for a sub-domain ${\cal D'}$ of ${\cal D}$ with (units are in K):
\begin{equation}
\begin{aligned}
\tilde{{\sf J}}_{z} = &-0.5 \pm 0.05 \\
0.66 \approx &(1.5-\tilde{{\sf J}}_{x})(1.5-\tilde{{\sf J}}_{y})
\end{aligned}
\end{equation}
but one is left with a quite large uncertainty regarding $\tilde{{\sf J}}_{x}$ and $\tilde{{\sf J}}_{y}$. We note that the best $\chi^2$ is obtained for $\tilde{{\sf J}}_{x}\sim 0.16,~ \tilde{{\sf J}}_{y}\sim 0.97$ and $\theta = 69^{\circ}$, yet this choice does not predict the correct neutron intensities. \\

To solve this problem and resolve this uncertainty, we investigate more closely the sets of parameters found in ${\cal D'}$. To this end, we consider, for each set of coupling constants in ${\cal D'}$, a trial $\theta$ such that $0 \leq \theta \leq \pi/2$, and compare the calculated and measured magnetic moments for the different ions in the unit cell (see Supplementary Note 1 which discusses the neutron diffraction). A new $\chi^2$ is defined as the sum of squared differences between those quantities for a series of fields between -1 and 1 T. For: 
\begin{equation}
\tilde{{\sf J}}_{z} =( -0.5 \pm 0.05)~{\rm K},~
\tilde{{\sf J}}_{x} = (1.0 \pm 0.05)~{\rm K},~
\tilde{{\sf J}}_{y} = (0.066 \pm 0.2)~\mbox{K}, 
\theta = 72 \pm 10^{\circ}
\end{equation}
 corresponding to
 \begin{equation}
{\sf J}_{x} = (-0.36 \pm 0.16)~{\rm K},~
{\sf J}_{y} = (0.066 \pm 0.2)~{\rm K},~
{\sf J}_{z} = (0.86 \pm 0.15)~{\rm K},~
{\sf J}_{xz} = (0.44 \pm 0.15)~{\rm K}
\end{equation}
we find a good agreement with INS results and the field induced magnetic structure. Note that the latter values seem to strongly differ from those given in Supplementary Reference \citenum{Benton16b}, but this difference is essentially due to the value of $\theta$. The dispersions along $(-2,h,2-h)$ shown in Supplementary Figure \ref{calc1}(a) at $\mu_0H=0.25$ and 0.75 T capture the rise of the B branch, as well as the shift and the narrowing of the A branch with increasing field. Calculated constant energy maps at $H=0$ and 0.25 T are also shown in Supplementary Figures \ref{calc2} and \ref{calc3}. The energies are chosen to describe the flat modes and the two (A and B) dispersing branches already observed in Supplementary Figure \ref{calc1}. The calculations show that both of them stem from the pinch points and close at the $(-2,2,0)$ wavevectors, as in experiment. Importantly, the structure factor of the flat mode strongly evolves with field. At $\mu_0H=0.25$ T, the calculation especially demonstrates the rise of a dynamical kagome ice pinch points pattern (left upper panel of Supplementary Figure \ref{calc3}). \\ 

Furthermore, the calculated magnetic structure experiences a field evolution which resembles the diffraction data. Supplementary Figures \ref{calc1}(b) and (c) compare the calculated magnetic moment of the apical \nd\ ion ($y$) and of the three kagome ions ($x$) with diffraction data, at 50 mK and 0.5 K respectively. While the overall agreement is good, there is a significant discrepancy between the experimentally observed field at which the abrupt transition occurs and the field value produced by the calculation at 50 mK. In the experiment, it is due to the transition from the all in -- all out configuration to the 3 in -- 1 out structure and occurs at about 0.08 T. In the calculation, a similar abrupt transition also occurs, but at about 1~T at 50 mK and 0.35 T at 0.5 K. The origin of this transition is better understood when looking at the $\tau_x$ components of the pseudo-spin. Supplementary Figure \ref{calc1}(d) shows that the transition occurs when the $\tau_x$ component of the apical spins changes sign, becoming positive. In other words, it corresponds to the field where the apical spins are fully aligned along the ${\bf z}$ axis. Because of the primarily antiferromagnetic $\tilde{\sf J}_z$ interactions, an all in -- all out configuration is always favoured. Here the field already imposes a 3 in -- 1 out configuration for the ${\bf z}$ components, which is not favourable. A gain in exchange energy can be however achieved if the ${\bf x}$ components keep the same sign, hence precipitating the transition.

\vspace{-0.3cm}
\begin{figure*}[h!]
\includegraphics[height=6cm]{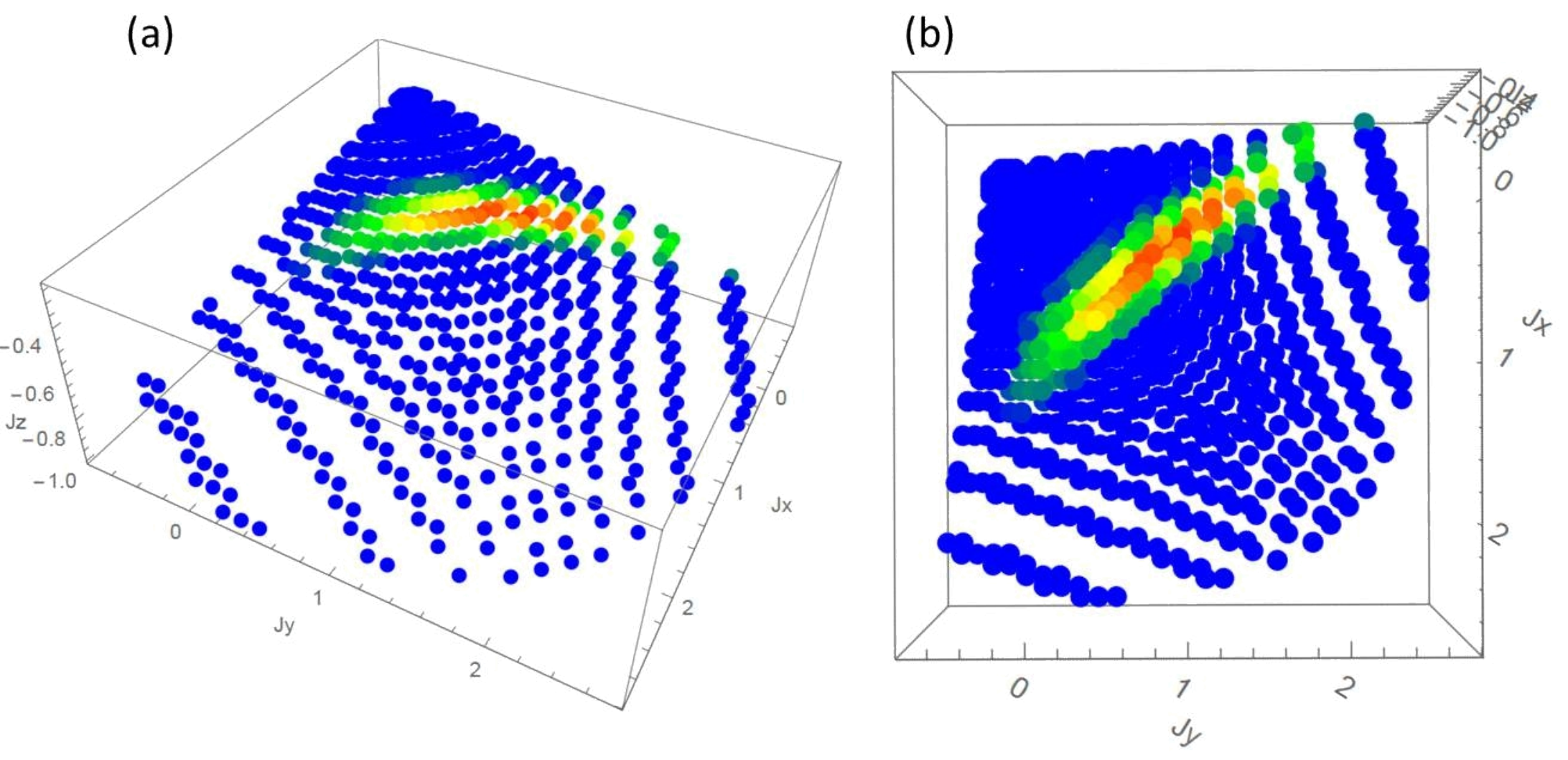}
\caption{{\bf Inverse of $\chi^2$ as a function of the Hamiltonian parameters.} The colour-map is chosen such that the best values are in red and the worst ones in blue. (a) Map as a function of the three parameters $\tilde{{\sf J}}_{x}, \tilde{{\sf J}}_{y}$ and $\tilde{{\sf J}}_{z}$. (b) Same data projected onto the $(\tilde{{\sf J}}_{x}, \tilde{{\sf J}}_{y}$) plane.}
\label{chi2}
\end{figure*}


\begin{figure*}[h]
\includegraphics[width=\textwidth]{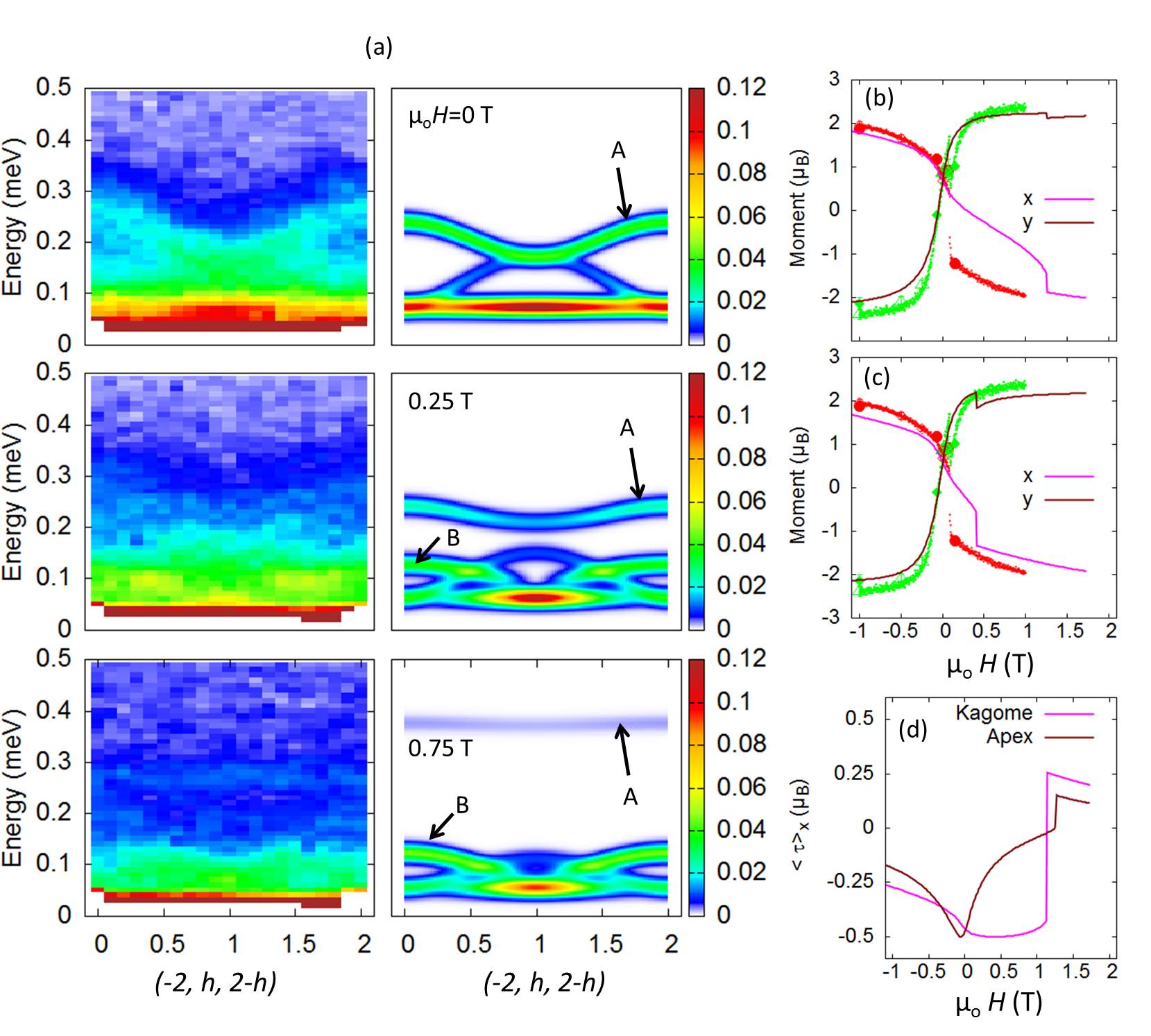}
\caption{{\bf Comparison between measurements and calculations for the dispersions and the ordered moment. }(a) Measured and calculated dispersions along $(-2,h,2-h)$ for 0, 0.25 and 0.75 T. The parameters are ${\sf J}_{x} = -0.36~{\rm K}, {\sf J}_{y} = 0.066~{\rm K}, {\sf J}_{z} = 0.86~{\rm K}, {\sf J}_{xz} = 0.44$ K. Above 0.25 T, a new dispersive branch (labelled B) rises at low energy. The original branch (labelled A) already present in zero field narrows and shifts towards higher energy with increasing field. {\bf (b)} and {\bf (c)} Calculated magnetic moment of the apical \nd\ ion ($y$) and of the three kagome ions ($x$) at 50 mK and 0.5~K respectively. The transition is clearly better reproduced at the effective temperature of 0.5 K. {\bf (d)} Component $\tau_x$ of the pseudo-spin. Interestingly, it is always of the same sign for both the apical and kagome pseudo-spins. This results from the large negative value of $\tilde{{\sf J}}_{z}$. }
\label{calc1}
\end{figure*}


\begin{figure*}[h]
\includegraphics[width=16 cm]{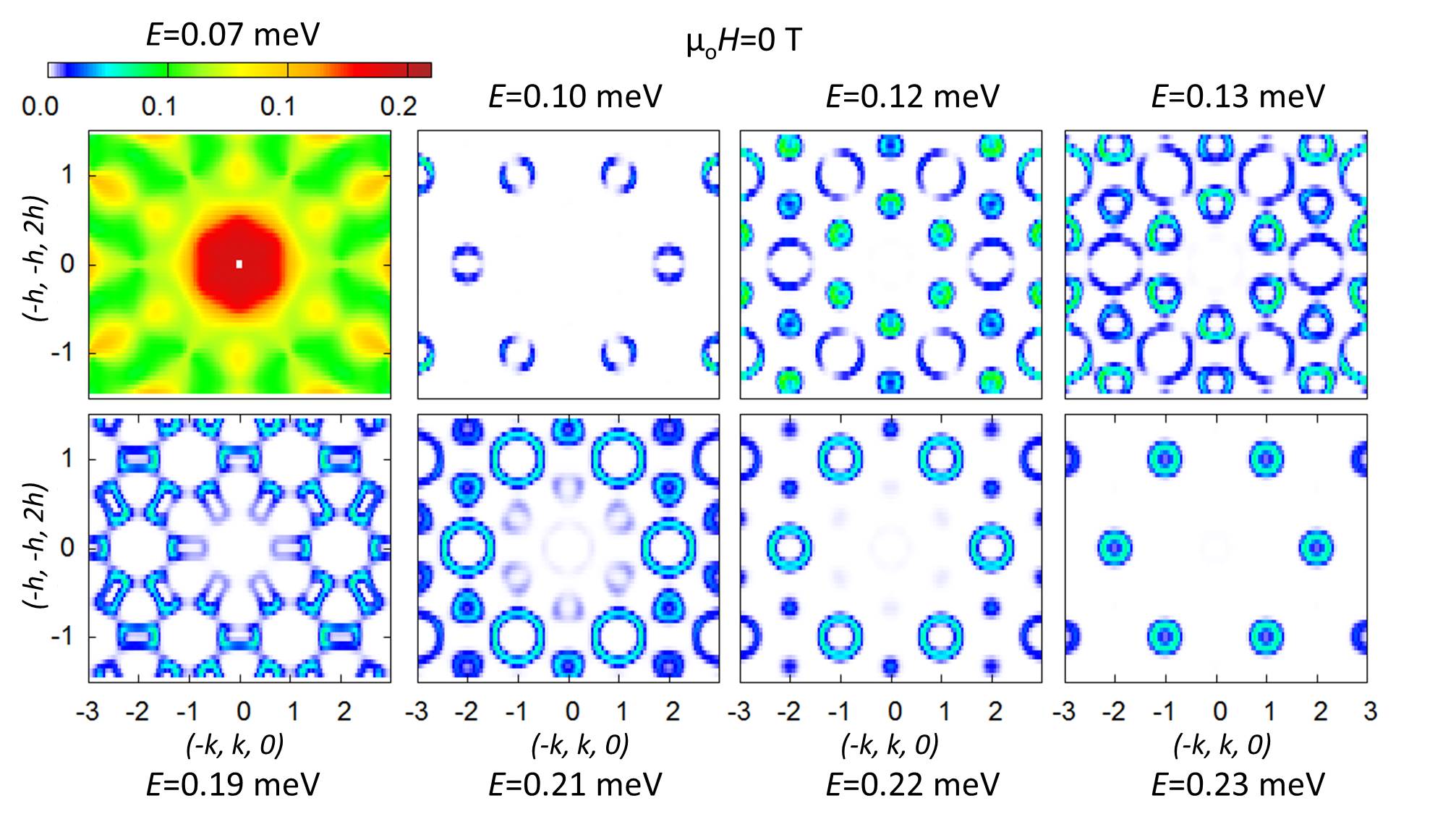}
\caption{{\bf Calculated constant energy maps at zero field.} The energies are chosen to describe the flat and dispersing modes that can be also observed in Supplementary Figure \ref{calc1}.}
\label{calc2}
\end{figure*}


\begin{figure*}[h]
\includegraphics[width=16 cm]{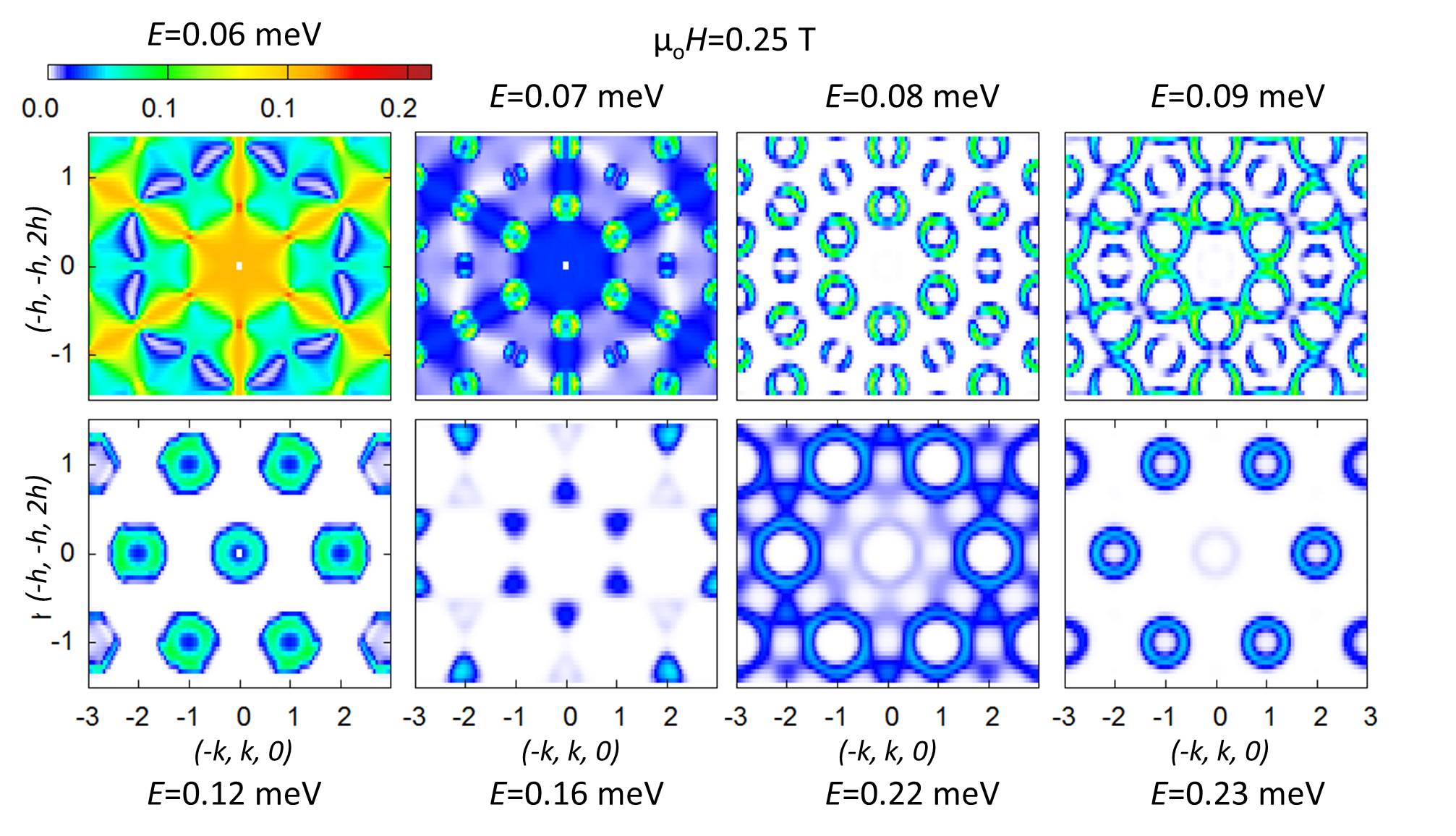}
\caption{{\bf Calculated constant energy maps at 0.25 T}. The energies are chosen to describe the new dispersing modes as well as the higher energy branch. Comparing the flat mode structure factor with the one displayed in Supplementary Figure \ref{calc2} clearly demonstrates the rise of pinch points in the scattering plane.}
\label{calc3}
\end{figure*}

\clearpage 
\subsection{Decomposition of the spectrum into divergence free and divergence full dynamical fields}

The decomposition of the spectrum in terms of (lattice) divergence free and divergence full dynamical fields has first been shown in Supplementary Reference \citenum{Benton16b}. Here, we follow a similar route, expanding the calculations by taking into account the effect of the magnetic field. We first write the equations of motion in zero field taking advantage of the fact that the XYZ model is written using site dependent local bases:
\begin{equation}
\frac{d {\bf m}_i}{dt} = {\bf m}_i \times \sum_j \bar{{\sf J}}~ {\bf m}_j
\end{equation}
where the sum over $j$ sites is restricted to nearest neighbours and $\bar{{\sf J}}$ is the exchange tensor. Here, ${\bf m}_i$ is written in the local basis attached to site $i$, and not in the global Cartesian frame. \\

Our derivation assumes that the ground state is uniform, i.e. that the spins locally point along the same {\it local} direction, which is nothing but an all in -- all out like configuration. In other terms, we assume here that the system is essentially an antiferromagnet. Calling ${\bf m}^{\rm o}$ the local equilibrium magnetization, we linearize those equations to obtain:
\begin{equation}
\frac{d \delta {\bf m}_i}{dt} = {\bf m}^{\rm o} \times \sum_j \bar{{\sf J}}~ \delta{\bf m}_j + \delta{\bf m}_i \times 6~\bar{{\sf J}}~ {\bf m}^{\rm o}.
\end{equation}
Note that ${\bf m}^{\rm o}$ is indeed identical, whatever the site, since it is written in local bases. \\

Since ${\bf m}^{\rm o}$ is an equilibrium solution, we also note that:
\begin{equation}
0 = {\bf m}^{\rm o} \times \sum_j \bar{{\sf J}}~ {\bf m}^{\rm o}
\end{equation}
hence, ${\bf m}^{\rm o}$ and $\bar{{\sf J}}~{\bf m}^{\rm o}$ must be collinear. Writing the exchange tensor in the diagonal form (see section III): 
\begin{equation}
\bar{{\sf J}} = 
\tilde{{\sf J}}_{x}~
\tilde{{\bf x}} \tilde{{\bf x}}+
\tilde{{\sf J}}_{y}~
\tilde{{\bf y}} \tilde{{\bf y}}+
\tilde{{\sf J}}_{z}~
\tilde{{\bf z}} \tilde{{\bf z}}
\end{equation}
we observe that if ${\bf m}^{\rm o}$ is along one of these axes, for instance $\tilde{{\bf z}}$, then 
\begin{equation}
\bar{{\sf J}}~ {\bf m}^{\rm o}=
\tilde{{\sf J}}_{z}~
(\tilde{{\bf z}}\cdot{\bf m}^{\rm o})~\tilde{{\bf z}} = 
\tilde{{\sf J}}_{z}~m^{\rm o}~\tilde{{\bf z}}.
\end{equation}
In other words, ${\bf m}^{\rm o}$ and $\bar{{\sf J}}~{\bf m}^{\rm o}$ are indeed collinear. The energy is finally minimized when ${\bf m}^{\rm o}$ is along the axis corresponding to the smallest eigenvalue among $(\tilde{{\sf J}}_{x},\tilde{{\sf J}}_{y},\tilde{{\sf J}}_{z})$.

Thanks to the pyrochlore structure, the sum over the neighbours sites $j$ can be written in terms of the components of the generalized charge: 
\begin{equation}
Q_{\boxtimes}^{u=x,y,z} = \sum_{i \in \boxtimes} m_{i}^u
\end{equation} 
defined in each tetrahedron $\boxtimes$ (see Supplementary Figure \ref{Fig6}). Note that by definition, those charges reside on the dual lattice of the pyrochlore, which is the diamond lattice. Since the latter is bipartite, we shall consider two dual sub-lattices, $A$ and $B$, with: 
\begin{equation}
Q_{\boxtimes A}^u = +\sum_{i \in \boxtimes A} m_{i}^u
\end{equation} 
for $A$ diamond sites and 
\begin{equation}
Q_{\boxtimes B}^u = -\sum_{i \in \boxtimes B} m_{i}^u
\end{equation}
for $B$ diamond sites. Interestingly, the uniform static solution ${\bf m}^{\rm o}$ can be understood as a staggered pattern, from the point of view of the charge, with ${\bf Q}_{\boxtimes} = \pm4{\bf m}^{\rm o}$. Finally, using these definitions of ${\bf Q}_{\boxtimes {A,B}}$, we shall write
the equation of motion as:
\begin{equation}
\begin{aligned}
\frac{d \delta{\bf m}_i}{dt}=& {\bf m}^{\rm o} \times \bar{{\sf J}}~ 
\left(
{\bf Q}_{\boxtimes_i A} - \delta{\bf m}_i - ({\bf Q}_{\boxtimes_i B} + \delta{\bf m}_i)
\right) + 
\delta{\bf m}_i \times 6~\bar{{\sf J}}~ {\bf m}^{\rm o} \\
=& {\bf m}^{\rm o} \times \bar{{\sf J}}~ 
\left(
{\bf Q}_{\boxtimes_i A} - {\bf Q}_{\boxtimes_i B} -2 \delta{\bf m}_i
\right) + 
\delta{\bf m}_i \times 6~\bar{{\sf J}}~ {\bf m}^{\rm o}
\end{aligned}
\end{equation}
where the indexes $\boxtimes_i A$ and $\boxtimes_i B$ denote the $A$ and $B$ tetrahedra the site $i$ belongs to. This form of the linearized equations of motion is the starting point to derive a peculiar partition of the solutions. 
\bigskip

{\bf Charged solutions: }
We start with the subset of charged solutions. Forming the equations of motion for ${\bf Q}_{\boxtimes {A,B}}$ using the previous equations, one obtains:
\begin{equation}
\begin{aligned}
\frac{d {\bf Q}_{\boxtimes A}}{dt} =& {\bf m}^{\rm o} \times \bar{{\sf J}}~ \left(4~{\bf Q}_{\boxtimes A} - \sum_{\boxtimes B}{\bf Q}_{\boxtimes B} - 2~{\bf Q}_{\boxtimes A}\right) + {\bf Q}_{\boxtimes A} \times 6~\bar{{\sf J}}~ {\bf m}^{\rm o} \\
=& {\bf Q}_{\boxtimes A} \times 6~\bar{{\sf J}}~ {\bf m}^{\rm o} + {\bf m}^{\rm o} \times 2~\bar{{\sf J}}~ {\bf Q}_{\boxtimes A} - \sum_{\boxtimes B} {\bf m}^{\rm o} \times \bar{{\sf J}}~ {\bf Q}_{\boxtimes B}.
\end{aligned}
\end{equation}
Those solutions correspond to the propagation of the charge throughout the diamond lattice. We shall also make several points:
\begin{itemize}
\item the $A \leftrightarrow B$ symmetry is clear from the above equation.
\item The sum over $\boxtimes B$ refers to the four neighbouring $B$ tetrahedra of a given $A$ tetrahedron. 
\item Since for a lattice containing $N$ site, there are $(N/4+N/4)$ $A$ and $B$ tetrahedra, the number of these modes is $N/2$. In the Helmholtz-Hodge decomposition of the ${\bf m}_i$ field, these modes are obviously associated with the dynamical divergence full field.
\item Finally, we write the exchange tensor in its diagonal form $
{\bar{\sf J}} = 
\tilde{{\sf J}}_{x}~
{\tilde{\bf x}} {\tilde{\bf x}}+
\tilde{{\sf J}}_{y}~
{\tilde{\bf y}} {\tilde{\bf y}}+
\tilde{{\sf J}}_{z}~
{\tilde{\bf z}} {\tilde{\bf z}}$ (assuming that ${\bf m}^{\rm o}$ is along $\tilde{\bf z}$, while $(\tilde{\bf x}, \tilde{\bf y})$ span the plane perpendicular to ${\bf m}^{\rm o}$) and introduce the coordinates of the charge in the corresponding ($\tilde{\bf x}, \tilde{\bf y}, \tilde{\bf z}$) frame:
\begin{equation}
{\bf Q}_{\boxtimes A} = 
\left(
\begin{array}{c}
Q^{\tilde{x}}_{\boxtimes A} \\
Q^{\tilde{y}}_{\boxtimes A} \\
Q^{\tilde{z}}_{\boxtimes A} \\
\end{array}
\right)
\end{equation}
The $Q^{\tilde{z}}_{\boxtimes A}$ components remain constant (consistent with the all in -- all out structure), while the $Q^{\tilde{x}}_{\boxtimes A}$ and $Q^{\tilde{y}}_{\boxtimes A}$ components evolve in time following:
\begin{equation}
\frac{d {\bf Q}_{\boxtimes A}}{dt} 
= 
\left(
\begin{array}{cc}
0 & 6 \tilde{{\sf J}}_{z} m^{\rm o} - 2 \tilde{{\sf J}}_{y} m^{\rm o} \\
-(6 \tilde{{\sf J}}_{z} m^{\rm o} - 2 \tilde{{\sf J}}_{x} m^{\rm o}) & 0
\end{array}
\right)
{\bf Q}_{\boxtimes A} ~-~ 
\sum_{\boxtimes B} 
\left(
\begin{array}{cc}
0 & \tilde{{\sf J}}_{y} m^{\rm o} \\
- \tilde{{\sf J}}_{x} m^{\rm o} & 0
\end{array}
\right)~ {\bf Q}_{\boxtimes B}.
\end{equation}
\end{itemize}
\bigskip

{\bf Alternate loops weathervane modes: }
The second set of solutions is constructed by introducing loops ${\cal L}$ consisting of $N$ sites in the pyrochlore lattice. Note that the shortest of these loops are the hexagons that form the kagome planes. We consider:
\begin{equation}
{\bf A}=\sum_{k \in {\cal L}} u_k \delta{\bf m}_{k}.
\end{equation}
Forming the equation of motion for ${\bf A}$, we obtain:
\begin{equation}
\begin{aligned}
\frac{d {\bf A}}{dt} = & {\bf A} \times 6~\bar{{\sf J}}~ {\bf m}^{\rm o} - {\bf m}^{\rm o} \times 2~\bar{{\sf J}}~{\bf A} \\
 &+{\bf m}^{\rm o} \times \bar{{\sf J}}~
\left(
u_1 {\bf Q}_{\boxtimes_1A} - u_1{\bf Q}_{\boxtimes_1B} + ... + u_N {\bf Q}_{\boxtimes_NA} - u_N{\bf Q}_{\boxtimes_NB}
\right).
\end{aligned}
\end{equation}
We note that two successive sites belong, for example, to the same $A$ tetrahedron but to different $B$ ones: 
$\boxtimes_1A = \boxtimes_2A,~\boxtimes_3A = \boxtimes_4A,~ ... = ... ,~\boxtimes_{N-1}A = \boxtimes_NA$
and
$\boxtimes_2B = \boxtimes_3B,~\boxtimes_4B = \boxtimes_5B,~ ... = ...$\\
Furthermore, since ${\cal L}$ is a loop, we have $\boxtimes_NB= \boxtimes_1B$, hence:
\begin{equation}
\begin{aligned}
\sum_{k=1,N}
u_k {\bf Q}_{\boxtimes_kA} =&
(u_1+u_2){\bf Q}_{\boxtimes_1A} + (u_3+u_4){\bf Q}_{\boxtimes_3A} + ... \\
\sum_{k=1,N}
u_k {\bf Q}_{\boxtimes_kB} =& (u_N+u_1) {\bf Q}_{\boxtimes_1B} + (u_2+u_3) {\bf Q}_{\boxtimes_1B} + ... 
\end{aligned}
\end{equation}
Assuming alternate coefficients $u_{k+1}=-u_k$, and $u_N=-u_1$, we obtain: 
\begin{equation}
\frac{d {\bf A}}{dt} = {\bf A} \times 6~\bar{{\sf J}}~{\bf m}^{\rm o} - {\bf m}^{\rm o} \times 2~\bar{{\sf J}}~{\bf A}
\end{equation}
This shows that the alternate loop solutions correspond to a resonating dispersionless field, decoupled from the charged solutions. To determine the corresponding energy, we write the exchange tensor in its diagonal form: ${\bar{\sf J}} = \tilde{{\sf J}}_{x}~{\tilde{\bf x}} {\tilde{\bf x}}+\tilde{{\sf J}}_{y}~{\tilde{\bf y}} {\tilde{\bf y}}+\tilde{{\sf J}}_{z}~{\tilde{\bf z}} {\tilde{\bf z}}$, and assume that $\tilde{\bf z}$ is along ${\bf m}^{\rm o}$. We thus have:
\begin{equation}
\begin{aligned}
\frac{d {\bf A}}{dt} =& 
6 \tilde{{\sf J}}_{z} m^{\rm o} 
( A_{\tilde{x}} {\tilde{\bf x}} + A_{\tilde{y}} {\tilde{\bf y}}) \times {\tilde{\bf z}} - 
2 m^{\rm o} {\tilde{\bf z}} \times ( 
\tilde{{\sf J}}_{x} {\tilde{\bf x}} A_{\tilde{x}} +
\tilde{{\sf J}}_{y} {\tilde{\bf y}} A_{\tilde{y}} +
\tilde{{\sf J}}_{z} {\tilde{\bf z}} A_{\tilde{z}}) \\
=& 6 \tilde{{\sf J}}_{z} m^{\rm o} ( 
- A_{\tilde{x}} {\tilde{\bf y}} 
+ A_{\tilde{y}} {\tilde{\bf x}}) 
- 2 m^{\rm o} ( 
\tilde{{\sf J}}_{x} {\tilde{\bf y}} A_{\tilde{x}} - \tilde{{\sf J}}_{y} {\tilde{\bf x}} A_{\tilde{y}} )\\
=&
\left(
\begin{array}{cc}
0 & 6 \tilde{{\sf J}}_{z} m^{\rm o} + 2 \tilde{{\sf J}}_{y} m^{\rm o} \\
-(6 \tilde{{\sf J}}_{z} m^{\rm o} + 2 \tilde{{\sf J}}_{x} m^{\rm o}) & 0
\end{array}
\right)
{\bf A}
\end{aligned}
\end{equation}
The eigenvalues of above matrix are of the form $\pm i \Delta$ with:
\begin{equation}
\Delta= 2 m^{\rm o}~\sqrt{(3 \tilde{{\sf J}}_{z}+ \tilde{{\sf J}}_{x})(3 \tilde{{\sf J}}_{z}+ \tilde{{\sf J}}_{y})}
\end{equation}
and $\Delta$ corresponds to the frequency of those dispersionless modes. Since we consider only the cases where ${\tilde{\sf J}}_{z}<0$, we write ${\tilde{\sf J}}_{z}=-|{\tilde{\sf J}}_{z}|$ and finally:
\begin{equation}
\Delta= 2 m^{\rm o}~\sqrt{(3 |\tilde{{\sf J}}_{z}| - \tilde{{\sf J}}_{x})(3 |\tilde{{\sf J}}_{z}| - \tilde{{\sf J}}_{y})}
\end{equation}
which is the energy of the flat mode described in Supplementary notes 3 and 4.
\bigskip

{\bf Influence of a $[111]$ magnetic field: }
Applying a magnetic field along one of the $\langle 111 \rangle$ high symmetry directions is especially interesting as the sites of the pyrochlore lattice separate in kagome ($K$) and triangular ($T$) planes forming the subset of apical spins. The latter are rapidly polarized by the field since the Ising direction is precisely along ${\bf H}$. The equations of motion can now be written as:
\begin{equation}
\frac{d {\bf m}_i}{dt} = {\bf m}_i \times \left( \sum_j \bar{{\sf J}}~ {\bf m}_j + g_i {\bf H} \right)
\end{equation}
and we still consider a linearized version around a uniform solution. However, as emphasized above, the field has different influences depending on the site: for the apical sites, $g_i {\bf H} = - g_{\parallel} H~{\bf z}$ while for the kagome sites, $g_i {\bf H} = g_ {\parallel} H/3~{\bf z}$. As a result, we shall consider a generalized solution ${\bf m}^{o,(K,T)}$ with different equilibrium values for the apical and kagome sites:
\begin{equation}
\begin{aligned}
0 =& {\bf m}^{o,T} \times \left( 6~\bar{{\sf J}}~ {\bf m}^{o,K} - g_ {\parallel} H~{\bf z} \right) \\
0 =& {\bf m}^{o,K} \times \left( 4~\bar{{\sf J}}~ {\bf m}^{o,K} + 2 ~\bar{{\sf J}}~ {\bf m}^{o,T} + g_ {\parallel} H/3~{\bf z} \right).
\end{aligned}
\end{equation}

Our aim is to show that once the apical spins are fully polarized, the kagome ice displays, on one hand charged solutions, and on the other, flat alternate loop modes. Owing to the dimensional reduction, the charges are, however, kagome charges, and the loops are confined within the kagome layers. To this end, we introduce the kagome charge ${\bf Q}_{\bigtriangleup}^{u=x,y,z} = \sum_{i \in \bigtriangleup} m_{i}^u$ of a given kagome triangle $\bigtriangleup$. It is worth noting that the same bipartite property holds in this 2D case. As a result, $A$ and $B$ triangles can be defined, in close analogy with the pyrochlore case (see Supplementary Figure \ref{Fig7}). Let us consider, on one hand, a site $i$ where the $i$ index runs in the kagome plane. 
This site belongs to an $A$ and to a $B$ kagome triangle, and has two neighbours sites in the $T$ triangular planes, one in the top ($j=t_i$) plane and one in the bottom ($j=b_i$) plane. The equation of motion can be written as:
\begin{equation}
\frac{d \delta{\bf m}^K_i}{dt} = 
{\bf m}^{o,K} 
\times \bar{{\sf J}} 
\left( 
{\bf Q}_{\bigtriangleup_i A} - \delta{\bf m}^K_i - \left( {\bf Q}_{\bigtriangleup_i B} + \delta{\bf m}^K_i \right)+\sum_{j=t_i,b_i} \bar{{\sf J}}~\delta{\bf m}^T_j 
\right)
+ \delta{\bf m}^K_i \times 
\left( 
4 \bar{{\sf J}}~{\bf m}^{o,K}+2 \bar{{\sf J}}~ {\bf m}^{o,T} + g_{\parallel} H/3~{\bf z}\right).
\end{equation}
On the other hand, let's consider a site $j$ where $j$ runs in the triangular plane. This site is also the apical site of $A$ and $B$ kagome triangles, denoted $\bigtriangleup_j A$ and $\bigtriangleup_i B$. With this notation, the equation of motion becomes:
\begin{equation}
\frac{d \delta{\bf m}^T_j}{dt} = 
{\bf m}^{o,T} \times 
\bar{{\sf J}}~ 
\left(
{\bf Q}_{\bigtriangleup_j A} - {\bf Q}_{\bigtriangleup_j B}
\right) 
+ \delta{\bf m}^T_j \times 6~\bar{{\sf J}}~ {\bf m}^{o,K} 
- g_{\parallel} H~\delta{\bf m}^T_j \times {\bf z}.
\end{equation}
If the field is strong enough to fully polarize the apical sites, we may neglect the term $\sum_{j=t_i,b_i} \bar{{\sf J}}~\delta{\bf m}^T_j $, and hence we obtain:
\begin{equation}
\frac{d {\bf m}^K_i}{dt} = 
{\bf m}^{o,K} 
\times \bar{{\sf J}} 
\left( 
{\bf Q}_{\bigtriangleup_i A} - {\bf Q}_{\bigtriangleup_i B} - 2 \delta{\bf m}^K_i 
\right) 
+ \delta{\bf m}^K_i \times 
\left( 
4 \bar{{\sf J}}~ {\bf m}^{o,K}+2 \bar{{\sf J}}~ {\bf m}^{o,T} + g_{\parallel} H/3~ {\bf z}
\right).
\end{equation}
In this limit, we obtain a decomposition of the excitations similar to what we found in zero field, in terms of divergence full and divergence free fields, but within the kagome layers. The charged solutions are given by:
\begin{equation}
\frac{d {\bf Q}_{\bigtriangleup A}}{dt} = 
{\bf Q}_{\bigtriangleup A} \times 
\left( 
4 \bar{{\sf J}}~ {\bf m}^{o,K}+2 \bar{{\sf J}}~ {\bf m}^{o,T} + g_{\parallel} H/3 ~ {\bf z}
\right) 
+
{\bf m}^{o,K} 
\times \bar{{\sf J}}~{\bf Q}_{\bigtriangleup A}
- \sum_{\bigtriangleup B}
{\bf m}^{o,K} 
\times \bar{{\sf J}}~{\bf Q}_{\bigtriangleup B}
\end{equation}
and the dispersionless solutions by hexagonal loops of alternate spins: 
\begin{equation}
\frac{d {\bf A}}{dt} = 
{\bf A} \times 
\left( 
4 \bar{{\sf J}}~ {\bf m}^{o,K}+2 \bar{{\sf J}}~ {\bf m}^{o,T} + g_{\parallel} H/3~ {\bf z}\right) 
-~{\bf m}^{o,K} 
\times \bar{{\sf J}}~{\bf A}.
\end{equation}


\begin{figure*}[h]
\includegraphics[height=8cm]{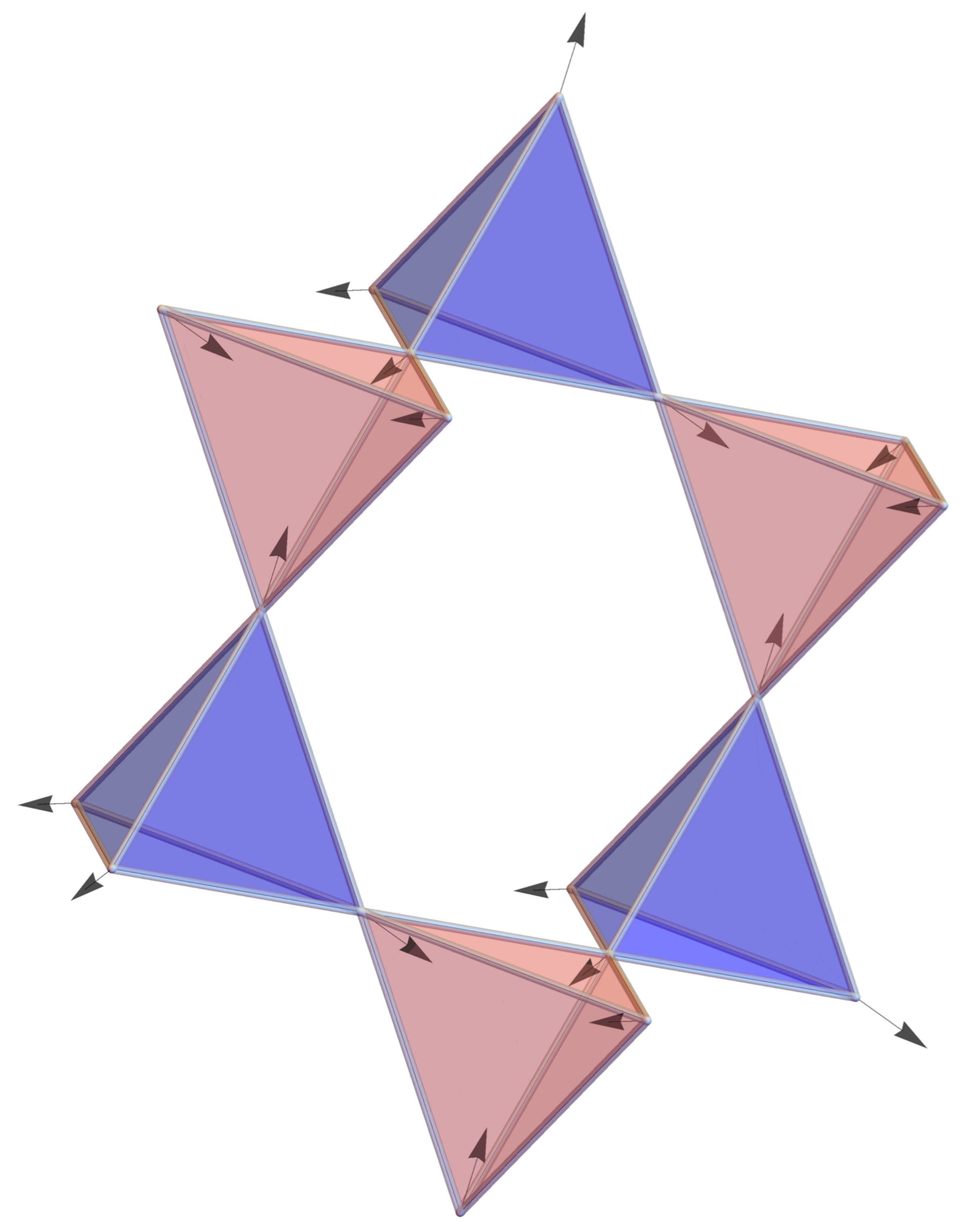}
\caption{{\bf Partial view of the pyrochlore lattice in the all in -- all out magnetic structure depicted by the black arrows}. The staggered charged pattern is shown using blue and red tetrahedra. The nodes of the diamond lattice, dual of the pyrochlore, are the centres of the tetrahedra. This dual lattice hosts the charge. It is bipartite, which allows one to distinguish $A$ and $B$ charges. Clearly, each pyrochlore site belongs to an $A$ and to a $B$ tetrahedron.}
\label{Fig6}
\end{figure*}


\begin{figure*}[h]
\includegraphics[width=10cm]{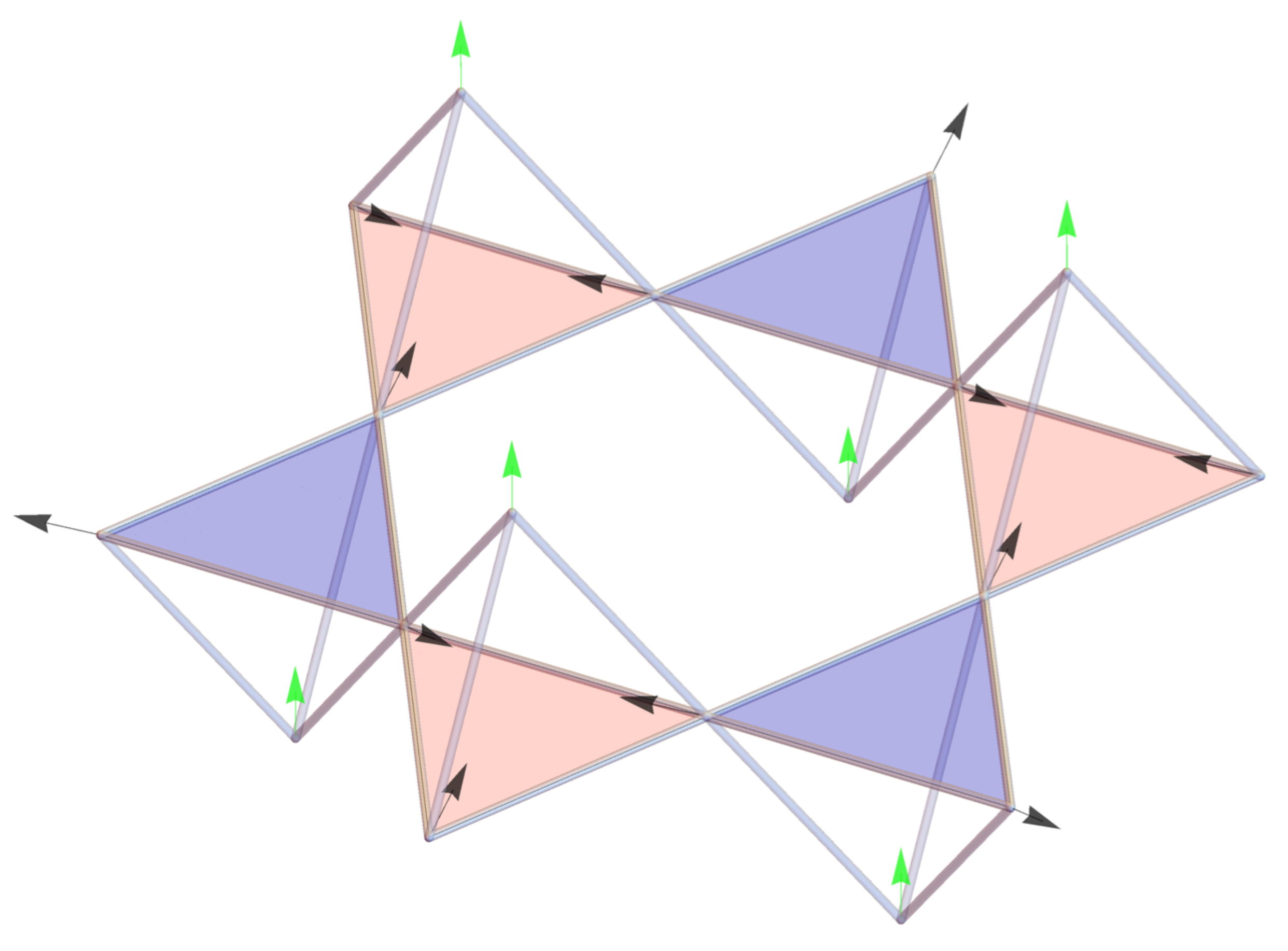}
\caption{{\bf Partial view of the pyrochlore lattice in the 3 in -- 1 out magnetic structure}. The black and the green arrows depict the kagome and apical spins respectively. The staggered charged pattern corresponding to a uniform solution is shown using blue and red triangles.}
\label{Fig7}
\end{figure*}

\vspace{2cm}
~\\


\end{document}